\documentclass[a4paper,12pt]{article}
\usepackage{graphicx}
\usepackage{amsmath, latexsym}
\numberwithin{equation}{section}
\newcommand{\be}{\begin{equation}}
\newcommand{\eeq}{\end{equation}}
\newcommand{\emt}{\end{math}}
\newcommand{\bmt}{\begin{math}}
\newcommand{\bdm}{\begin{displaymath}}
\newcommand{\edm}{\end{displaymath}}
\newcommand{\bes}{\begin{equation} \begin{split}}
\newcommand{\ees}{\end{split} \end{equation}}
\newcommand{\bea}{\begin{eqnarray}}
\newcommand{\eea}{\end{eqnarray}}

\allowdisplaybreaks

\author{Ghasem Exirifard\footnote{exir@sissa.it}, Martin O'Loughlin\footnote{loughlin@sissa.it}}

\begin{document}

\rightline{SISSA/61/2004/EP}
\vfill

\begin{center}{\Large 
Two and three loop $\alpha'$ corrections to T-duality:\\
Kasner and Schwarzschild. }
\end{center}

\begin{center}{\large 
Ghasem Exirifard\footnote{e-mail: {\tt exir@sissa.it}},
Martin O'Loughlin\footnote{e-mail: {\tt loughlin@sissa.it}}}
\end{center}

\vskip 0.05 cm
\centerline{\it  ISAS -- SISSA, Via Beirut 2-4, I-34013 Trieste, Italy}

\vskip 0.05 cm
\centerline{\it  INFN, Sezione di Trieste, Italy}

\begin{abstract}
Two and three loop $\alpha'$ corrections are calculated for Kasner and 
Schwarzschild metrics, and their T-duals, in the bosonic string theory. 
These metrics are 
used to obtain the two and three loop $\alpha'$ corrections to T-duality. 
It is noted in particular that the inclusion of alpha' corrections and the
requirement of consistency with the alpha'-corrected T-duality for the
Kasner and Schwarzschild metrics enables one to fix uniquely the 
covariant form of the T-duality rules at three loops. 
As a generalization of the T-dual of the 
Schwarzschild geometry a class of massless geometries is presented. 

\vspace{0.5cm}
\vspace{6cm}

\end{abstract}

\pagebreak



\section{Introduction and Motivation}
Target space duality was first introduced as a symmetry describing the interchange of the momentum and winding modes in the closed string compactified on a torus\cite{firsttduality}. Later it was described as the symmetry of sigma models\cite{buscher}.  
The linear $\alpha'$-corrections to the T-duality rules are obtained in \cite{Tseylinduality,olsen,HOS}. Further support for the $\alpha'$ expansion of T-duality 
is presented in \cite{kaloper} where the linear $\alpha'$ corrections 
to T-duality in the presence of torsion is obtained.
Studying higher $\alpha'$ corrections to T-duality should provide a better understanding of both the mathematics of string theory in the curved space time and the pre-big bang scenario in string cosmology \cite{ven} where T-duality is an essential tool.

In this paper the three loop $\alpha'$ corrections to T-duality are computed in the critical bosonic string theory at the tree level of the string interaction for backgrounds composed of diagonal metric and dilaton. The paper is organized in the following way:

In the second section we are going to review the general diagonal Kasner backgrounds in $D=26$. Since the Kasner background is of interest in Cosmology\cite{BKL1,BKL2} and particularly in Cosmological Billiard \cite{Henneaux} calculating its string corrections should be  interesting. We also present the 
$\beta$-function equations of the bosonic string theory.

In the third section we generalize the  Kasner metric  to a perturbative background in the critical bosonic string and we calculate the linear (two-loop) and  the quadratic (three-loop) $\alpha'$ corrections  to this background at the tree level of the string interaction.

In the fourth section we write the Kasner metric on a periodic space-like directions and we apply T-duality in one direction to obtain the corresponding T-dual background. Next we add the linear and the quadratic $\alpha'$ corrections to the Kasner background and to its T-dual. We will observe that T-duality fails to relate the $\alpha'$-corrected Kasner background to its $\alpha'$ corrected T-dual background. We will modify the rules by appropriate $\alpha'$ terms in such a way that the $\alpha'$ modified rules relate the $\alpha'$-corrected Kasner background to its $\alpha'$ corrected T-dual background. Finally we will rewrite the $\alpha'$-modifications in a Lorentz invariant form consistent with \cite{olsen} to obtain the $\alpha'$ corrected  T-duality rules for a general time-dependent background with diagonal metric and dilaton.

In the fifth section we review the Schwarzschild background in an arbitrary dimension. We introduce the time-dual of the Schwarzschild background  by performing T-duality in the time direction of the related Euclidean geometry. We observe that the horizon of the Schwarzschild background changes into an intrinsic singularity under T-duality and the time-dual of the Schwarzschild metric is massless in $D=4$. We introduce massless geometries in arbitrary dimensions for the low energy gravitational theory of bosonic string theory.  
We then calculate the linear and the quadratic $\alpha'$ corrections to both the Schwarzschild background and its T-dual background in $D=4$ in the critical bosonic string theory at the tree-level of the string interaction. We observe that when the asymptotic behaviours of the fields are fixed at infinity the $\alpha'$ corrections generically diverge at the horizon of the Schwarzschild black hole.The possible relevance of this divergence is discussed.

Finally, by requiring that the $\alpha'$ modified T-duality relate the quadratic $\alpha'$ corrected Schwarzschild background to the quadratic $\alpha'$ corrected time-dual, we are able to identify uniquely the quadratic $\alpha'$ correction to the rule  which describes the change of the dilaton under T-duality.

In the last section we provide a summary and possible generalizations of this work.  

\section{Kasner metrics and $\alpha'$ corrections}
Exact conformal symmetry in string theory requires  vanishing of the $\beta$ functions of the corresponding sigma model. The vanishing of $\beta$-functions gives the equations of motion which describe the dynamics of the corresponding low energy theory. 
We will consider only the case of the critical bosonic string theory where the leading  equations for the background composed of the metric and  dilaton in the string frame read
\begin{eqnarray}\label{eq1action}
0&=&R_{ij}~+~2~\bigtriangledown_i\bigtriangledown_j~\phi~, \\
0&=& \Box \phi~-~(\bigtriangledown\phi)^2~+~\frac{1}{4}~R~. \label{eq2action} 
\end{eqnarray}
The stringy Kasner background \cite{kasner} 
determined by the set of $\{p_1,\cdots,p_{_{25}}\}$ as a solution of the above equations is
\begin{eqnarray}
ds^2&=&~-~dt^2~+~\sum_{i=1}^{25} t^{2p_i} ~ dx_i^2 \label{casnerat26}~,\\
\phi(t)&=&\frac{\sum p~-~1}{2}~\ln t\nonumber~,\\
&& \sum_{i=1}^{25}~p_i^2~=~1\label{cons}~,
\end{eqnarray}
where satisfying the equations imposes the constraint (\ref{cons}). The intrinsic singularity at $t=0$

of this metric shows up in the various scalar curvature terms  
\begin{eqnarray}
R&=&
~\frac{(\sum p~-~1)^2}{t^2}~,\\\nonumber\\
R_{\mu\nu}~R^{\mu\nu}&=&
~2~\frac{(\sum p~-~1)^2}{t^4}~,
\\\nonumber\\
R_{\mu\nu\lambda\eta}~R^{\mu\nu\lambda\eta}&=&~
~\frac{6~+~2~\sum p^4~-~8~\sum p}{t^4}~,\\
\nonumber\\
\bigtriangledown_{\xi}\,R_{\mu\nu\lambda\eta}~\bigtriangledown^{\xi}\,R^{\mu\nu\lambda\eta}&=&16~\frac{-\sum p^4~+~2\sum p^3~+~(\sum p^3)^2~-~2}{t^6}~,
\end{eqnarray}
where $R$, $R_{\mu\nu}$ and $R_{\mu\nu\lambda\eta}$ stand for the Ricci scalar, Ricci and Riemann tensors respectively.

The usual Kasner metric has $\phi\,=\,0$ as $\sum p\,-\,1\,=\,0$.\footnote{The subleading $\beta$ function (\ref{phi2loop}) provides a non-zero value for the dilaton even if it vanishes at the leading order} Here we allow for a more general configuration with a time dependent string coupling.  The string coupling constant given by the local vacuum expectation value of the dilaton   reads
\begin{equation}
g_s~=~g_0~e^{\,\phi}~=~g_0~t^{(\sum p\,-\,1\,)/2}~.
\end{equation}
Therefore for positive values of $\sum p\, -\, 1$ the string coupling constant vanishes at the time origin  and diverges at infinity. For negative values of 
$\sum p\, -\, 1$ the string coupling constant diverges at the time origin  and vanishes at infinity.

In this work we are interested in calculating the $\alpha'$ corrections at the tree level of the string interaction. The calculation at the tree level can be trusted as long as $g_s\ll1$. For negative values of $\sum p -1$ this condition is automatically satisfied at large $t$ and for positive values of $\sum p -1$ we set $g_0$ close to $0$ to get $g_s\ll 1$ at the vicinity of a fixed large value of time where the perturbation in $\alpha'$ is going to be done.

In general one can calculate higher-loop corrections to the $\beta$-functions and obtain the $\alpha'$-corrections to the equations of motion. Note that generally the $\beta$-functions are scheme dependent 
 and the $\beta$-functions of various schemes 
should be
 mapped to each other by an appropriate field redefinition.

The $\beta$-functions of the cirital bosonic string theory for the backgrounds of dilaton and metric, calculated by the dimensional regularization method in the minimal substraction scheme at three loops in $\alpha'$ read\cite{Jack,Tsytlin},
\begin{eqnarray}\label{g2loop}
\frac{1}{\alpha'}
\beta_{ij}&=&R_{ij}~+~2~\bigtriangledown_i\bigtriangledown_j~\phi~+~\frac{1}{2}~\alpha'~R_{iklm}R_j^{~~klm}\\
&&+\alpha'^2\left\{\frac{1}{8}\bigtriangledown_k R_{ilmn}\bigtriangledown^k R_{j}^{~~lmn}-\frac{1}{16}\bigtriangledown_iR_{klmn}\bigtriangledown_jR^{klmn}+\frac{1}{2}R_{klmn}~R_{i}^{~~mlp}~R_{j~~~p}^{~~kn}
\right.\nonumber\\
&&\qquad~~~~~\left.~-~\frac{3}{8}~R_{iklj}~R^{kmnp}~R^{l}_{~~mnp}~+~\frac{1}{32}~\bigtriangledown_j\bigtriangledown_i(~R_{klmn}~R^{klmn})\right\}=~0~,\nonumber\\
\frac{1}{\alpha'}
\beta_\phi&=&~-~\frac{1}{2}~\Box~\phi~+~\partial_k\phi~\partial^k \phi~+~\frac{1}{16}~\alpha'~R_{klmn}~R^{klmn}\label{phi2loop}\\
&&~+~\alpha'^2~\left\{-~\frac{3}{16}~R^{kmnp}~R^{l}_{~~mnp}~\bigtriangledown_k\bigtriangledown_l\phi~+~\frac{1}{32}~R_{klmn}~R^{mnpq}~R_{pq}^{~~~kl}\right.\nonumber\\
&&\left.\qquad~-~\frac{1}{24}~R_{klmn}~R^{qnpl}~R_{q~~p}^{~m~~k}~+~\frac{1}{64}~\partial_i(R_{klmn}~R^{klmn})~\partial^i\phi\right\}~=~0~,\nonumber\\
&&i,\cdots,q~\in~\{0...25\}\nonumber~.
\end{eqnarray}
The $\alpha'$ terms in (\ref{g2loop}) and (\ref{phi2loop}) are the string modifications to the metric and dilaton.  In the following sections we refer to these equations to compute the $\alpha'$ corrections to the general Kasner background and the Schwarzschild metric in $D=4$.

\allowdisplaybreaks
\section{Perturbative $\alpha'$ corrections to the Kasner metric}
In the previous section, the Kasner background was defined as the solution to the leading order $\beta$-function equations. Before moving to the subleading order corrections, one needs to generalize the Kasner background to  a perturbative   background in string theory. We implement this generalization by requiring that
\begin{enumerate}\label{bgassume}
\item String theory admits some time-dependent backgrounds where the metric is globally diagonal and the only non-vanishing field is  the dilaton.
\item The above backgrounds admit a perturbative series expansion in $\alpha'$ i.e.
\begin{eqnarray}
g_{\mu\nu}(t)&=&g_{\mu\nu}^{(0)}(t)~+~\alpha'~g_{\mu\nu}^{(1)}(t)~+~\alpha'^2~g_{\mu\nu}^{(2)}(t)~+~\cdots~,\\
\phi(t)&=&\phi^{(0)}(t)~+~\alpha'~\phi^{(1)}(t)~+~\alpha'^2~\phi^{(2)}(t)~+~\cdots~,
\end{eqnarray}
where  $g_{\mu\nu}^{(0)}(t)$ and $\phi^{(0)}(t)$ correspond to the Kasner background. All $g_{\mu\nu}^{(n)}$'s become automatically diagonal due to the first assumption.
\end{enumerate}  
We expect that for every given Kasner background  there exists a string background satisfying the above conditions. In the next sections we are going to compute the linear and the quadratic $\alpha'$ corrections to the general Kasner background.

\subsection{The linear $\alpha'$ correction to the Kasner metric}
We begin to investigate the $\alpha'$ corrections to the Kasner metric by 
making the following simple ansatz the generality of which we will verify at the
end of this subsection. 
\begin{eqnarray}
ds^2~&=&~-~dt^2~+~\sum_{i=1}^{25}~t^{2p_i}~(1~+~2\frac{\alpha'}{t^2}~b_i)~dx_i^2~+~O(\frac{\alpha'^2}{t^4})~,\label{metricfirst}\\ 
\phi(t)~&=&~-~\frac{1}{2}(1-\sum_{i=1}^{25}~p_i)~\ln t~+~\frac{\alpha'~B}{2~t^2}~+~O(\frac{\alpha'^2}{t^4})~,\label{dilatonfirst}\\
&&\sum_{i=1}^{25}~p_i^2~=~1~,\nonumber
\end{eqnarray}
where $b_i$'s and $B$ are some unknown constants numbers. 
Substituting (\ref{metricfirst}) and (\ref{dilatonfirst}) in  (\ref{g2loop}) and keeping only the linear $\alpha'$ term results in the algebraic equations, 
\begin{eqnarray}
2~b_i~+~(-\sum_{j=1}^{25}~b_j~+~B)~p_i~+~p_i^2~-~p_i^3 ~=~0~, \label{one}\\
6~(-\sum_{i=1}^{25}~b_i~+~B)~-~\sum_{i=1}^{25}~p_i^4~-~1~+~2~\sum_{i=1}^{25}~p_i^3~+~4~\sum_{i=1}^{25}~p_i~b_i~=~0\label{two}~.
\end{eqnarray}
Multiplying (\ref{one}) by $p_i$ and summing over $i$ gives
\begin{equation}\label{three}
2~\sum_{i=1}^{25}~b_i~p_i~+~(-\sum_{i=1}^{25}~b_i~+~B)~+~\sum_{i=1}^{25}~p_i^3~-~\sum_{i=1}^{25}~p_i^4~=~0~.
\end{equation}
(\ref{three}) and (\ref{two}) are solved by 
\begin{equation}\label{Bminusb}
B~-~\sum_{i=1}^{25}b_i~=~\frac{1}{4}(1~-~\sum_{i=1}^{25} p^4)~.
\end{equation}
 Using (\ref{one}) and (\ref{Bminusb}) one easily obtains
\begin{eqnarray}
b_i~&=&~-~p_i~\left(\frac{1~-~\sum_{j=1}^{25}~p_j^4}{8}~+~\frac{1}{2}(p_i~-~p_i^2)\right)~,\label{generalcasnerbfirstalpha}\\
B~&=&~(1~-~\sum_{j=1}^{25}~p_j^4)~(\frac{1}{4}~-~\frac{\sum_{i=1}^{25}~p_i}{8})~-~\frac{1}{2}(1~-~\sum_{i=1}^{25}~p_i^3)~.
\label{generalcasnerfirstalphaphi}
\end{eqnarray}
The  results obtained for $b_i$ and $B$ satisfies (\ref{phi2loop}) as well. Now let us investigate the general solution by writing  the corrections in the following form
\begin{eqnarray}
ds^2&=&~-~dt^2~+~\sum_{i=1}^{25} t^{2 p_i}~(1~+~\frac{2\alpha'}{t^2}~(b_i(t)~+~b_i))~dx_i^2~+~O(\frac{\alpha'^2}{t^4})~,\label{tbfirst}\\
\phi(t)&=&\frac{1}{2}~(\sum p~-~1)~\ln t~+~(B~+~B(t))~\frac{\alpha'}{2~t^2}~+~O(\frac{\alpha'^2}{t^4})\label{tphifirst}~,
\end{eqnarray}
where $b_i$ and $B$ are given respectively in (\ref{generalcasnerbfirstalpha}) and (\ref{generalcasnerfirstalphaphi}). In order to find $b_i(t)$ and $B(t)$ we first define the following variables 
\begin{eqnarray}
x(t)&=& B(t)~-~\sum_{i=1}^{25} b_i(t)~,\\
y(t)&=& \sum_{i=1}^{25} b_i(t)~p_i~.
\end{eqnarray}
Inserting (\ref{tbfirst}) and (\ref{tphifirst}) in (\ref{g2loop}) and (\ref{phi2loop}) and keeping only the linear term in $\alpha'$ yields\footnote{(\ref{Modifieddilaton}) is obtained from $\frac{1}{4}~g^{ij}~\beta_{ij}~-~\beta_\phi$}
\begin{eqnarray}
-~2~x(t)~+~\frac{3}{2}~x'(t)~t~-~\frac{1}{2}~x''(t)~t^2~-~y(t)~+~\frac{1}{2}~y'(t)~t &=& 0~,\label{Modifieddilaton}\\
6~x(t)~-~4~t~x'(t)~+~x''(t)~t^2~+~4~y(t)~-~2~y'(t)~t &=& 0~,\\
2~b_i(t)~-~\frac{3}{2}~t~b'_i(t)+~\frac{t^2}{2}~b''_i(t)~+~p_i~(x(t)~-~\frac{t}{2}~x'(t))&=&0~.
\end{eqnarray}
The general solution of the above system is  
\begin{eqnarray}
b_i(t)&=&-~p_i~c_1~t~+~c^{(1)}_i~t^2~+~2~t^2~c^{(2)}_i~\ln t \label{arbitaryconstantb}~,\\
B(t)&=&c_1~(1~-~\sum p)~t~+~c_2~t^2~+~2~t^2~\ln t~\sum c^{(2)}_i \label{arbitaryconstantB}~,\\
\sum c^{(2)}_i~p_i&=&0 ~\label{arbitarycons}~,
\end{eqnarray}
where $c_1, c_2$, $c^1_i$'s and $c^2_i$'s are constants of integration. At the first sight the appearance of these constants of integration may seem disappointing, however a closer look shows that 
\begin{itemize}
\item $c_1$ corresponds to an infinitesimal time displacement  , $t~\to ~t~-~\alpha'~c_1$.
\item $c_2$ corresponds to a constant shift in the dilaton field.
\item $c^{(1)}_i$'s correspond to proper scaling in the $x_i$ directions.
\item $c^{(2)}_i$ describes an infinitesimal change in $p_i$, $p_i\to p_i+2\alpha'c_i^{(2)}$, constrained to $\sum p_i^2=1$. (\ref{arbitarycons}).
\end{itemize}
Therefore all the arbitrary constants in (\ref{arbitaryconstantb}) and (\ref{arbitaryconstantB}) are infinitesimal redefinitions of the variables.  We fix the definition of the variables and set all of the arbitrary constants to zero. Doing so we obtain  (\ref{generalcasnerbfirstalpha}) and (\ref{generalcasnerfirstalphaphi}) as the values of the linear $\alpha'$ corrections to  the metric and the dilaton.\footnote{In the Heterotic string theory the $\beta$ functions at two-loop in $\alpha'$ are the same as the ones of bosonic string theory by replacing $\alpha'$ with $\frac{\alpha'}{2}$. Therefore the results obtained in   (\ref{generalcasnerbfirstalpha}) and (\ref{generalcasnerfirstalphaphi}) trivially can be extended to Heterotic string theory.}

\subsection{The quadratic $\alpha'$ correction to the Kasner metric}
Similar to  what was done in the previous section the quadratic $\alpha'$ correction to the metric and the dilaton may be written as 
\begin{eqnarray}
ds^2&=&-dt^2~+~\sum_{i=1}^{25}~t^{2p_i}~(1~+~2~\frac{\alpha'}{t^2}~b_i~+~2~\frac{\alpha'^2}{t^4}~c_i)~+~O(\frac{\alpha'^3}{t^6})~,\\
\phi(t)&=&\frac{\sum p-1}{2}~\ln t~+~\frac{\alpha'}{2~t^2}~B~+~\frac{\alpha'^2}{t^4}~W~+~O(\frac{\alpha'^3}{t^6})~, 
\end{eqnarray}
where $c_i$'s and $W$ are some unknown constant numbers and $b_i$'s and $B$ are written explicitly in the (\ref{generalcasnerbfirstalpha}) and (\ref{generalcasnerfirstalphaphi}). The equation that comes from the time-time component of (\ref{g2loop}) 
reads 
\begin{eqnarray}\label{eq11cas}
0&=&-\frac{17}{4}\sum p^3~\sum p^4~-\sum p^5~\sum p^4~-~\frac{\sum p^3~(\sum p^4)^2}{8}
\\
&&+\sum(-~\frac{45}{8}~p^3~+~\frac{17}{4}~p^4~-~19~p^5~+~11~p^6~-~2~p^7~+~14~p^4)\nonumber\\
&&+\frac{11}{4}~+~8\sum c~p+~20~(2~W~-\sum c)~.\nonumber
\end{eqnarray}
The remaining equations generated by (\ref{g2loop}) are 
\begin{eqnarray}\label{eqiicas}
0&=&\frac{3}{16}p_i~-~\frac{1}{4}~p_i^2~+~7~p_i^3~-~18~p_i^4~+~16~p_i^5~-~4~p_i^6 \\
&&+~p_i~\sum(p^6~-~2~p^5~+~\frac{5}{2}~p^3~-~\frac{15}{8}~p^4)~+~\frac{11}{16}~p_i~(\sum p^4)^2~\nonumber\\
&&-p_i^2~(~\frac{7}{2}~\sum p^4~+~\frac{1}{4}(\sum p^4)^2)~+~5~p_i^3~\sum p^4~-~2~p_i^4~\sum p^4\nonumber\\
&&-~\frac{p_i}{2}~\sum p^3~\sum p^5+~4~(2~W~-~\sum c)~p_i~+~16~c_i~~.\nonumber
\end{eqnarray}
Multiplying (\ref{eqiicas}) with $p_i$ and summing over $i$ gives
\begin{eqnarray}\label{eqtrcas}
0&=&-~4~\sum p^3~\sum p^4~-~2~\sum p^5~\sum p^4~-\frac{1}{4}\sum p^3~(\sum p^4)^2\\
&&~+~\sum (~-~20~p^5~+~17~p^6~+~\frac{9}{4}~p^3~+~\frac{41}{8}~p^4~-~4~p^7)\nonumber\\
&&~+~\frac{3}{16}~+~\frac{91}{16} (\sum p^4)^2~+~16~\sum c~p~+~4~(2~W~-~\sum c)\nonumber~.
\end{eqnarray}
From (\ref{eqiicas}) and (\ref{eqtrcas}) one obtains 
\begin{eqnarray}\label{WminusC}
2~W~-~\sum c~&=& \sum (\frac{3}{8}~p^3~-~\frac{61}{96}~p^4~+~\frac{1}{2}~p^5~-~\frac{5}{36}~p^6)~+~\\
&&~+~\frac{1}{8}~\sum~p^4~\sum~p^3~-~\frac{5}{64}~(\sum p^4)^2~-~\frac{85}{576}\nonumber~.
\end{eqnarray}
Substituting (\ref{WminusC}) in (\ref{eqiicas}) identifies the $c_i$ giving
\begin{eqnarray}
\label{csecond}
c_i&=&p_i~\left\{\frac{1}{4}~p_i^5~-~p_i^4~+~\frac{9}{8}~p_i^3~-~\frac{7}{16}~p_i^2~+~\frac{1}{64}~p_i~+~\frac{29}{1152}\right. \\
&&\qquad+~p_i~(\frac{7}{32}~+~\frac{1}{8}~p_i^2~-~\frac{5}{16}~p_i)~\sum p^4~+~\frac{1}{64}~p_i~(\sum p^4)^2\nonumber\\
&&\qquad\left.+~\sum(\frac{53}{192}~p^4~-~\frac{1}{36}~p^6)~-~\frac{3}{128}~(\sum p^4)^2~\underbrace{-~\frac{1}{4}~\sum p^3}\right\}~,\nonumber
\\\nonumber\\
\label{Wsecond}
W&=&-\frac{\sum p^3}{32}+\frac{17}{48}\sum p^4-\frac{\sum p^5}{4}+\frac{\sum p^6}{18}+\frac{1}{32}(\sum p^4)^2-\frac{3}{32}\sum p^3\sum p^4\nonumber\\
&&+\left\{\frac{53\sum p^4}{384}-\frac{\sum p^3}{8}-\frac{3(\sum p^4)^2}{256}+\frac{29}{2304}-\frac{\sum p^6}{72} \right\}\sum p-\frac{19}{288}~.
\end{eqnarray}
Having obtained $W$ and $c_i$ it is not difficult to find the general $\alpha'^2$ corrections to the metric and the dilaton. Let the quadratic $\alpha'$ corrections be written in the following way 
\begin{eqnarray}\label{time2casner}
ds^2&=&\,-\,dt^2\,+\,\sum t^{2p_i}\,\left(1\,+\,\frac{2~\alpha'}{t^2}\,b_i\,+\,\frac{2\,\alpha'^2}{t^4}\,(c_i\,+\,c_i(t)\right)\,dx_i^2\,+\,O(\frac{\alpha'^3}{t^6})~,\\
\phi(t)&=&~\frac{\sum p~-~1}{2}~\ln t~+~\frac{\alpha'}{2~t^2}~B~+~\frac{\alpha'^2}{t^4}~(W~+~W(t))
~+~O(\frac{\alpha'^3}{t^6})\nonumber~.
\end{eqnarray}
We get the following equations for the auxiliary variables $x(t)$ and $y(t)$ -defined below- as the result of substituting (\ref{time2casner}) in (\ref{g2loop}) and (\ref{phi2loop})\footnote{(\ref{Modifieddilaton2}) is obtained from $\frac{1}{4}~g^{ij}~\beta_{ij}~-~\beta_\phi$},
\begin{eqnarray}
x(t)~=~2~W(t)~-~\sum_{i=1}^{25}~c_i(t),\\
y(t)~=~\sum_{i=1}^{25} c_i(t)~p_i\qquad\qquad,\\
20~x(t)~-~8~x'(t)~t~+~x''(t)~t^2~-~2~y'(t)~t~+~8~y(t)~&=&0~,\\
16~c_i(t)~-~7~c_i'(t)~t~+~c_i''(t)~t^2~+~p_i~(4~x(t)~-~x'(t)~t)&=&0~,\\
-~8~x(t)~+~\frac{7}{2}~x'(t)~t~-~\frac{t^2}{2}~x''(t)~-~2~y(t)~+~\frac{t}{2}~y'(t)&=&0\label{Modifieddilaton2}~.
\end{eqnarray}
The general solution of the above ordinary system of differentiable equations in terms of the integration constants $c_1,c_2, c_i^{(1)}$ and $c^{(2)}_i$ is as follows 
\begin{eqnarray}
c_i(t)&=&~p_i~t^3~c_2~+~t^4~c^{(1)}_i~+~t^4~\ln t~c^{(2)}_i~,\\
W(t)&=&~c_1~t^4~+\frac{\sum p~-~1}{2}~t^3~c_2~+~t^4~\ln t~\frac{\sum c^{(2)}_i}{2}~,\\
\sum_{i=1}^{25} c^{(2)}_i~p_i&=&0 \nonumber~.
\end{eqnarray}
Again all of the above constants of integration can be eliminated by  redefining the variables appropriately. We set all of these constants of integration to zero and obtain (\ref{Wsecond}) and (\ref{csecond}) as the  quadratic $\alpha'$ corrections to the dilaton and the metric. Let us emphasize that setting the constants of integration to zero is the same as fixing the asymptotic behaviour of the metric and dilaton at infinity as we saw already at the end of the previous 
subsection. 

\section{T-Duality and the $\alpha'$ corrections}

In this section we are going to obtain the quadratic $\alpha'$ modifications to the T-duality rules for a time-dependent background composed of a 
diagonal metric and  dilaton, consistent with the results of the previous 
sections on the $\alpha'$ corrections to the Kasner metric and its T-dual.

We study T-duality in the effective field theory where it is a map from on-shell field contents of a given space to the  on-shell field contents of its corresponding T-dual space. In order to represent this map explicitly, we first represent both the field contents of a space ``$ds^2,\phi(t)$'' and the field contents of the corresponding  T-dual space ``$d\tilde{s}^2,\tilde\phi(t)$'' in the co-moving frame i.e.
\begin{equation}
\left\{
\begin{array}{rcl}
 ds^2&=&-\,dt^2\,+\,ds_\bot^2\\
 \phi&=&\phi(t)
\end{array}
\right.
\qquad,\qquad
\left\{
\begin{array}{rcl}
 d\tilde{s}^2&=&-\,dt^2\,+\,d\tilde{s}_\bot^2\\
 \tilde\phi&=&\tilde\phi(t)
\end{array}
\right.
\end{equation}
It should be noticed that the definition of the metric and the dilaton are always fixed in such a way  that the corresponding $\beta$-functions are provided by (\ref{g2loop}) and (\ref{phi2loop}).

In the following we write the Kasner metric on a periodic space-like direction and we apply T-duality in $x_{25}$ direction to obtain the corresponding T-dual background. Next we add the $\alpha'$ corrections to the Kasner background and to its T-dual. We will observe that T-duality fails to relate the $\alpha'$-corrected Kasner background to its $\alpha'$ corrected T-dual background. We will modify the rules by appropriate $\alpha'$ terms in such a way that the $\alpha'$ modified rules relate the $\alpha'$-corrected Kasner background to its T-dual.
 At the end of this section we will rewrite the $\alpha'$-modification in a Lorentz invariant form consistent with \cite{olsen} to obtain the $\alpha'$ corrected T-duality rules for a general time-dependent background with diagonal metric and dilaton.

Let us start the calculation by writing the Kasner background on periodic space directions 
\begin{eqnarray}
ds^2&=&-~dt^2~+~\sum_{i=1}^{25} t^{2p_i}~\left(\frac{r_i}{\sqrt{\alpha'}}\right)^2~dx_i^2~,\\
\phi(t)&=&\frac{\sum p~-~1}{2}~\ln t ~, \label{casnermet}\\
\sum_{i=1}^{25}p_i^2&=&1~,\\
x_i&\equiv&x_i~+~2~\pi~,
\end{eqnarray}
where each $x_i$ is compactified on a circle with time dependent radius $r_i(t)~=~r_i~t^{p_i}$.

 For the Kasner background the rules that describe T-duality in $x_{25}$ read \footnote{These rules are written in such a way that it is manifest that $T^2=1$, where T represents the T-duality}
\begin{eqnarray}\label{rules}
\ln \tilde{g}_{25~25}&=&-~\ln g_{25~25}~,\\
\ln \tilde{g}_{ii}&=&\ln g_{ii}~\quad~i\in\{1,\cdots, 24\}\label{tdrulep}~,\\
\tilde{\phi}~-~\frac{1}{4}~\ln{\det\tilde{g}}&=&{\phi}~-~\frac{1}{4}~\ln{\det{g}}\label{tdrulephi}~.
\end{eqnarray}
Applying the above rules on the Kasner background  returns its T-dual background,
\begin{eqnarray}\label{tdcasnermet}
d\tilde{s}^2&=&-~dt^2~+~\sum_{i=1}^{24} t^{2p_i}\left(\frac{r_i}{\sqrt{\alpha'}}\right)^2~d\tilde{x}_i^2+t^{-2p_{_{25}}}\left(\frac{r_{_{25}}}{\sqrt{\alpha'}}\right)^{-2}~d\tilde{x}_{_{25}}^2~,\\
\tilde\phi(t)&=&(\sum_{i=1}^{24}p_i\,-\,p_{_{25}}\,-\,1)~\frac{\ln t}{2}~,\\
\tilde{x}_i&\equiv&\tilde{x}_i~+~2~\pi \nonumber~.
\end{eqnarray}
The dual background is still  a Kasner background where $x_{25}$ is compactified on a circle with radius $\tilde{r}_{_{25}}(t)~=~\frac{\alpha'}{r_{_{25}}}~t^{^{-p_{_{25}}}}$ \cite{unpublished}.
Now let the $\alpha'$ corrections be added to the Kasner background 
\begin{eqnarray}
ds^2&=&-dt^2+\sum_{i=1}^{25}t^{2p_i}\left(\frac{r_i}{\sqrt{\alpha'}}\right)^2\{1\,+\,\frac{2\alpha'}{t^2}~b_i\,+\,\frac{2\alpha'^2}{t^4}c_i\}\,dx_i^2+O(\frac{\alpha'^3}{t^6})\label{metalpha}\\
\nonumber\\
\phi(t)&=&(\sum_{i=1}^{24}p_i\,+\,p_{_{25}}\,-\,1)~\frac{\ln t}{2}~+~\frac{\alpha'}{2\,t^2}B~+~\frac{\alpha'^2}{t^4}\,W~+~O(\frac{\alpha'^3}{t^6})~,
\end{eqnarray}
where $b_i$, $B$, $c_i$ and $W$ respectively are identified by (\ref{generalcasnerbfirstalpha}), 
(\ref{generalcasnerfirstalphaphi}), (\ref{csecond}) and (\ref{Wsecond}) for the set of $(p_1,\cdots,p_{_{24}},p_{_{25}})$.
Adding the corresponding $\alpha'$ corrections to the dual background gives 
\begin{eqnarray}
d\tilde{s}^2&=&-dt^2+\sum_{i=1}^{24}t^{2p_i}\left(\frac{r_i}{\sqrt{\alpha'}}\right)^2\{1\,+\,\frac{2\alpha'}{t^2}~\tilde{b}_i\,+\,\frac{2\alpha'^2}{t^4}\tilde{c}_i\}\,dx_i^2\label{mealphat}\\
&&\quad~+~t^{-2p_{_{25}}}\left(\frac{r_{_{25}}}{\sqrt{\alpha'}}\right)^{-2}\{1\,+\,\frac{2\alpha'}{t^2}~\tilde{b}_{_{25}}\,+\,\frac{2\alpha'^2}{t^4}\tilde{c}_{_{25}}\}\,dx_{_{25}}^2~,\\
\nonumber\\
\tilde{\phi}(t)&=&(\sum_{i=1}^{24}p_i\,-\,p_{_{25}}\,-\,1)~\frac{\ln t}{2}~+~\frac{\alpha'}{2\,t^2}\tilde{B}~+~\frac{\alpha'^2}{t^4}\,\tilde{W}~+~O(\frac{\alpha'^3}{t^6})~,
\end{eqnarray}
where $\tilde{b}_i$, $\tilde{B}$, $\tilde{c}_i$ and $\tilde{W}$ respectively are provided by (\ref{generalcasnerbfirstalpha}), (\ref{generalcasnerfirstalphaphi}), (\ref{csecond}) and (\ref{Wsecond}) for the set of $(p_1,\cdots,p_{_{24}},-p_{_{25}})$. It should be noticed that for $1\leq i \leq 24$ we have $b_i=\tilde{b}_i$ but due to the presence of the under-braced term in (\ref{csecond}) $c_i\neq\tilde{c}_i$.

Now let us check whether or not the T-duality rules (\ref{rules}, \ref{tdrulep}, \ref{tdrulephi}) map the $\alpha'$ corrected backgrounds (\ref{mealphat}, \ref{metalpha}) to each other. To perform this check it is better to write the rules describing the T-duality in the $x_{_{25}}$ direction in the following way
\begin{eqnarray}
\ln \tilde{g}_{_{25~25}}~+~\ln g_{_{25~25}}&=&0~,\\
\ln \tilde{g}_{ii}~-~\ln g_{ii}&=&0~,\\
\tilde{\phi}\,-\,\phi\,+\,\frac{1}{4}~(\,\ln \det g~-~\ln\det \tilde{g})&=&0~.
\end{eqnarray}
Substituting the $\alpha'$ corrected backgrounds in the l.h.s. of the above formulae gives
\begin{eqnarray}
\ln \tilde{g}_{_{25\,\,25}}+\ln g_{_{25\,\,25}}&=&\frac{2\alpha'}{t^2}\,(b_{_{25}}\,+\,\tilde{b}_{_{25}})+\frac{2\alpha'^2}{t^4}\,(c_{_{25}}+\tilde{c}_{_{25}}\,-\,b_{_{25}}^2-\tilde{b}_{_{25}}^2)+O(\frac{\alpha'^3}{t^6})~,\\
\nonumber\\
\ln \tilde{g}_{_{ii}}~-~\ln g_{_{ii}}&=&\frac{2\alpha'^2}{t^4}(c_i~-~\tilde{c}_i)~+~O(\frac{\alpha'^3}{t^6})~,\\
\nonumber\\
\tilde{\phi}-\phi+\frac{1}{4}\ln\frac{\det g}{\det \tilde{g}}&=&\frac{\alpha'}{2t^2}(\tilde{B}\,-\,B\,+\sum b-\sum \tilde{b})\\
&&+~\frac{\alpha'^2}{2t^2}(2\,\tilde{W}\,-\sum \tilde{c}+\sum \tilde{b}^2\,-\,2\,W\,+\sum c-\sum b^2)+O(\frac{\alpha'^3}{t^6})~.\nonumber
\end{eqnarray}
Keeping only the leading non-vanishing $\alpha'$ terms and expressing the r.h.s. of the above formulae in term of $p_1,\cdots,p_{_{25}}$ gives\footnote{Only the linear $\alpha'$ term in (\ref{mod1}) is kept. After finding the linear $\alpha'$ term in (\ref{tda1}) the quadratic term is fixed in (\ref{tda12}) .}  
\begin{eqnarray}
\ln \tilde{g}_{_{25\,\,25}}~+~\ln g_{_{25\,\,25}}&=&-\,\frac{2~\alpha'}{t^2}~p_{_{25}}^2~+~O(\frac{\alpha'^2}{t^4})\label{mod1}~,\\
\ln \tilde{g}_{ii}~-~\ln g_{ii}&=&~\frac{\alpha'^2}{t^4}~p_i~p_{_{25}}^3~+~O(\frac{\alpha'^3}{t^6})\label{mod2}~,\\\label{mod3}
\tilde{\phi}~-~\phi~+~\frac{1}{4}~\ln\frac{\det g}{\det \tilde{g}}&=&-\frac{\alpha'^2}{2~t^4}~p_{_{25}}^3~+~O(\frac{\alpha'^3}{t^6})
~.
\end{eqnarray}
\begin{figure}
   \centering
   \includegraphics[scale=0.7]{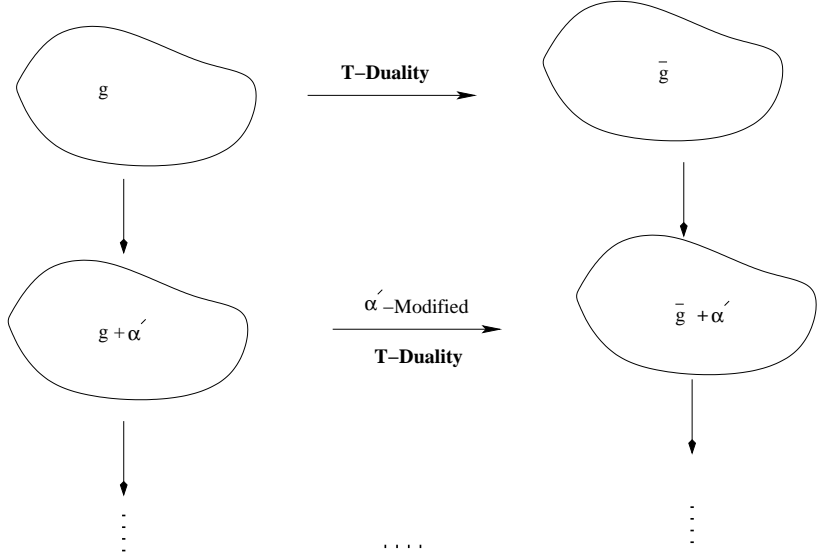}
   \caption{\textit{\small{\textbf{T-duality and the $\alpha'$ corrections}. By requiring that $\alpha'$ modified T-duality maps the $\alpha'$ corrected dual backgrounds to each other the $\alpha'$ modifications to the rules of the T-duality can be identified.}}}
   \label{talpha}
\end{figure}
The above relations indicate that none of the T-duality rules are satisfied and they all must be modified by appropriate $\alpha'$ terms (Fig.\ref{talpha}). In order to find the simplest modifications let (\ref{mod1}), (\ref{mod2}) and (\ref{mod3}) be written in the following forms 
\begin{eqnarray}
\ln \tilde{g}_{_{25\,25}}~+~\frac{\alpha'}{t^2}\tilde{p}_{_{25}}^2&=&~-~(\ln g_{_{25\,25}}~+~\frac{\alpha'}{t^2}p_{_{25}}^2)~+~O(\frac{\alpha'^2}{t^4})\label{td1}~,\\
\nonumber\\
\ln \tilde{g}_{_{ii}}~+~\frac{\alpha'^2}{2\,t^4}~\tilde{p}_i~\tilde{p}_{_{25}}^3&=&\ln g_{_{ii}}~+~\frac{\alpha'^2}{2\,t^4}~p_i~p_{_{25}}^3~+~O(\frac{\alpha'^3}{t^6})\label{td2}~,\\
\nonumber\\
\tilde{\phi}~-~\frac{1}{4}~\ln\det \tilde{g}~-~\frac{\alpha'^2}{4~t^4}~\tilde{p}_{_{25}}^3&=&\phi~-~\frac{1}{4}~\ln\det g~-~\frac{\alpha'^2}{4~t^4}~p_{_{25}}^3~+~O(\frac{\alpha'^3}{t^6})\label{td3}~,
\end{eqnarray}
where $~\tilde{p}_i=p_i~$ and $~\tilde{p}_{_{25}}=-p_{_{25}}~$. The simplest scheme to define the $\alpha'$ corrected T-duality is   to accept the above relations as the rules of the $\alpha'$-modified T-duality. We follow this strategy and we accept (\ref{td1}), (\ref{td2}) and (\ref{td3}) as the rules of the leading $\alpha'$ modified T-duality.

The leading $\alpha'$-modified T-duality explicitly depends on the $p_i$'s. These terms can be written as polynomials in derivatives of the metric and dilaton, in the following covariant forms\footnote{In this notation, the covariant derivative acts on the logarithm of the metric as if it were a scalar.}    
\begin{eqnarray}
\frac{p_i~p_{_{25}}^3}{t^4}&=&\frac{1}{16}~\nabla_\mu\ln g_{_{ii}}.\nabla^\mu\ln g_{_{25\,25}}~\nabla_\nu\ln g_{_{25\,25}}.\nabla^\nu\ln g_{_{25\,25}}~+~O(\alpha')~,\\\nonumber\\
\frac{p_{_{25}}^2}{t^2}&=&-\,\frac{1}{4}\nabla_\mu \ln g_{_{25\,25}}.\nabla^\mu\ln g_{_{25\,25}}~+~O(\alpha')~,\\\nonumber\\
-\frac{p_{_{25}}^3}{4~t^4}&=&\frac{A}{32}~\nabla_\mu\ln g_{_{25\,25}}~\nabla_\nu\ln g_{_{25\,25}}~\nabla^\mu\nabla^\nu\ln g_{_{25\,25}}\\
&&+~\frac{B}{16}~\nabla_\mu\ln g_{_{25\,25}}.\nabla^\mu\ln g_{_{25\,25}}~\nabla_\nu\ln g_{_{25\,25}}.\nabla^\nu(\phi-\frac{1}{4}\ln \det g)\nonumber\\
&&+~\frac{C}{32}~\nabla_\mu \ln g_{_{25\,25}}.\nabla^\mu\ln g_{_{25\,25}}~\Box\ln g_{_{25\,25}}~+~O(\alpha')\nonumber~,
\end{eqnarray}
where $A$, $B$ and $C$ are real numbers satisfying $A+B+C=1$. Using the above identities gives the leading $\alpha'$ modified  T-duality rules on the metric
\begin{eqnarray}\label{tda1}
\ln \tilde{g}_{_{25\,25}}&-&\frac{\alpha'}{4}\tilde{\nabla}_\mu\ln \tilde{g}_{_{25\,25}}.\tilde{\nabla}^\mu\ln \tilde{g}_{_{25\,25}}\\\nonumber
&=&-\left\{\ln {g}_{_{25\,25}}-\frac{\alpha'}{4}{\nabla}_\mu\ln {g}_{_{25\,25}}.{\nabla}^\mu\ln {g}_{_{25\,25}}\right\}+O(\alpha'^2\nabla^4)~,\\\nonumber\\\label{tda2}
\ln \tilde{g}_{_{ii}}&+&\frac{\alpha'^2}{32}\tilde{\nabla}_\mu\ln \tilde{g}_{_{25\,25}}.\tilde{\nabla}^\mu\ln \tilde{g}_{_{ii}}~\tilde{\nabla}_\nu\ln \tilde{g}_{_{25\,25}}.\tilde{\nabla}^\nu\ln \tilde{g}_{_{25\,25}}\\
&=&\ln {g}_{_{ii}}+\frac{\alpha'^2}{32}{\nabla}_\mu\ln {g}_{_{ii}}.{\nabla}^\mu\ln {g}_{_{25\,25}}~{\nabla}_\nu\ln {g}_{_{25\,25}}.{\nabla}^\nu\ln {g}_{_{25\,25}}+O(\alpha'^3\nabla^6)\nonumber~.
\end{eqnarray}
The T-duality rule which describes the change in the dilaton reads
\begin{eqnarray}\label{tda3}
\tilde{\phi}-\frac{1}{4}\ln \det\tilde g&+&\frac{\alpha'^2 A}{32}~\tilde{\nabla}_\mu\ln \tilde{g}_{_{25\,25}}~\tilde{\nabla}_\nu\ln \tilde{g}_{_{25\,25}}~\tilde{\nabla}^\mu\tilde{\nabla}^\nu\ln \tilde{g}_{_{25\,25}}\\
&+&\frac{\alpha'^2 B}{16}~\tilde{\nabla}_\mu\ln \tilde{g}_{_{25\,25}}.\tilde{\nabla}^\mu\ln \tilde{g}_{_{25\,25}}~\tilde{\nabla}_\nu\ln \tilde{g}_{_{25\,25}}.\tilde{\nabla}^\nu(\tilde{\phi}-\frac{1}{4}\ln \det\tilde g)\nonumber\\ 
&+&\frac{\alpha'^2 C}{32}~\tilde{\nabla}_\mu \ln \tilde{g}_{_{25\,25}}.\tilde{\nabla}^\mu\ln \tilde{g}_{_{25\,25}}~\tilde{\Box}\ln \tilde{g}_{_{25\,25}}\nonumber\\
\mathrm{=}~\phi&-&\frac{1}{4}\ln \det g~+~\frac{\alpha'^2 A}{32}~\nabla_\mu\ln g_{_{25\,25}}~\nabla_\nu\ln g_{_{25\,25}}~\nabla^\mu\nabla^\nu\ln g_{_{25\,25}}\nonumber\\
&+&\frac{\alpha'^2 B}{16}~\nabla_\mu\ln g_{_{25\,25}}.\nabla^\mu\ln g_{_{25\,25}}~\nabla_\nu\ln g_{_{25\,25}}.\nabla^\nu(\phi-\frac{1}{4}\ln \det g)\nonumber\\
&+&\frac{ \alpha'^2 C}{32}~\nabla_\mu \ln g_{_{25\,25}}.\nabla^\mu\ln g_{_{25\,25}}~\Box\ln g_{_{25\,25}}~+~O(\alpha'^3\nabla^6)\nonumber~,
\end{eqnarray}
where $A+B+C=1$. 

The above rules (\ref{tda1}), (\ref{tda2}) and (\ref{tda3}) are written in Lorentz invariant forms  and they describe  T-duality on backgrounds composed of diagonal metric and dilaton given that the fields are in the string frame and the co-moving frame . These rules are in agreement with those of  \cite{Tseylinduality} and \cite{olsen} where only the linear $\alpha'$ modifications were considered. 

One observes that redefining the metrics $g_{ii}$ and $\tilde g_{ii}$ to $g^*_{ii}$ and  $\tilde{g}^*_{ii}$ in the following way
\begin{eqnarray}
g^*_{ii}&=&g_{ii}\,\exp\left(\frac{\alpha'^2}{32}{\nabla}_\mu\ln {g}_{_{ii}}.\sum_{k=1}^{25}{\nabla}^\mu\ln {g}_{_{kk}}~{\nabla}_\nu\ln {g}_{_{kk}}.{\nabla}^\nu\ln {g}_{_{kk}}\right)~,\\
\tilde{g}^*_{ii}&=&\tilde{g}_{ii}\,\exp\left(\frac{\alpha'^2}{32}\tilde{\nabla}_\mu\ln \tilde{g}_{_{ii}}.\sum_{k=1}^{25}\tilde{\nabla}^\mu\ln \tilde{g}_{_{kk}}~\tilde{\nabla}_\nu\ln \tilde{g}_{_{kk}}.\tilde{\nabla}^\nu\ln \tilde{g}_{_{kk}}\right)~,
\end{eqnarray}
compensates the $\alpha'$ corrections to the rules describing  the change of the metric in the transverse directions under T-duality in any direction. For the 
Kasner metric this redefinition corresponds to choosing different coordinates on the space and its T-dual space. In general this transformation implies 
that one really needs a field redefinition to rewrite the higher order 
T-duality rules in the same form as the leading order T-duality rules.

Jack and Parson in \cite{odd} have calculated the same corrections to T-duality and proved that $O(d,d)$ invariance of the conformal  invariance condition, observed at one loop \cite{oddoneloop} and two loops \cite{oddtwoloop}, can be preserved  also at three loops given an appropriate field redefinition and coordinate transformation (either on the background or its T-dual background but not both). They did not explicitly provide the modification to T-duality but rewriting their results for the case of the Kasner background reproduces (\ref{tda2}).

In this work {\it we do not redefine the metric and the dilaton} and we maintain the convention given by the dimensional regularization method in the minimal substraction scheme which is the same as  fixing the definition of the metric and dilaton  in such a way that the corresponding $\beta$ functions are provided by (\ref{g2loop}) and (\ref{phi2loop}). This convention implies (\ref{tda2}) which means that applying T-duality in one direction alters the metric in all directions.

 Sometimes    writing the fields explicitly in the co-moving frame is not easy. Thus we are going to write the $\alpha'$ T-duality rules in an alternative frame. As a simple generalization of the co-moving frame let us introduce an {``\it almost co-moving frame'' } with a {\it`` characteristic function (f)''} in which  the metrics read
\begin{eqnarray}
ds^2\,=\,-\,f(T)\,dT^2\,+\,ds_\bot^2 &\quad,\quad&\phi\,=\,\phi(T)\label{fspace}\\
d\tilde{s}^2\,=\,-\,f(T)\,dT^2\,+\,d\tilde{s}_\bot^2 &\quad,\quad&\tilde\phi\,=\,\tilde\phi(T)\label{fspaced}
\end{eqnarray}
where $f(T)$ can be an arbitrary function of time. The $\alpha'$ modified T-duality rules in the almost co-moving frame for a general characteristic function are provided by those of the co-moving frame if within (\ref{tda3}) we replace $\det g$ and $\det\tilde g$ respectively by ${\det}^*g$ and ${\det}^*\tilde g $ given below
\begin{eqnarray}
{\det}^* g &=& \frac{\det g}{f(T)}~,\\
{\det}^* \tilde g &=& \frac{\det \tilde g}{f(T)}~.
\end{eqnarray}
For a specific background, one may choose  an appropriate characteristic function to simplify the compuations.

\section{The linear and the quadratic $\alpha'$ corrections to the \\
Schwarzschild background and its T-dual in $D=4$}

In this section the Schwarzschild background in an arbitrary dimension is reviewed. The time-dual of Schwarzschild background is introduced by performing a non-compact T-duality in the time direction of the Schwarzschild metric in the
region outside the black hole horizon. 
We must note the this non-compact T-duality in a time-like direction is not on the same footing as the usual T-duality, but it has been studied in \cite{Hullduals}. The linear and the quadratic $\alpha'$ corrections to the Schwarzschild background and its time-dual in $D=4$ are computed. Requiring (\ref{tda3}) to relate the quadratic $\alpha'$ corrected Schwarzschild background to the quadratic $\alpha'$ corrected its time-dual we are able to identify the unknown coefficients in (\ref{tda3}) with the values $A=1$ and $B=C=0$.

The Schwarzschild background in $D$ dimensions, ($D>3$), is given by
\begin{eqnarray}\label{schwarzschild}
ds^2&=&-\,(1\,-\,\frac{1}{r^{D-3}})\,dt^2\,+\,\frac{dr^2}{1\,-\,\frac{1}{r^{D-3}}}\,+\,r^2\,d\Omega_{D-2}~,\\
\phi(r)&=&0\nonumber~,
\end{eqnarray}
where the mass has been chosen to give simply a coefficient of one 
in the metric and we will maintain this convention in the following sections. 

The intrinsic singularity of the Schwarzschild metric shows itself in various scalar curvatures 
\begin{eqnarray}
R_{\mu\nu\lambda\eta} R^{\mu\nu\lambda\eta}&=&\frac{(D-1)\,(D-2)^2\,(D-3)}{r^{2D-2}}~,\\
\nabla_\xi R_{\mu\nu\lambda\eta}\,\nabla^\xi R^{\mu\nu\lambda\eta}
&=&\frac{(D\,+\,1)\,(D\,-\,1)^2\,(D\,-\,2)^2\,(D\,-\,3)}{r^{3\,D\,-\,3}}\,(r^{D-3}\,-\,1)~.
\end{eqnarray} 

Applying T-duality to the time direction, on the metric outside the 
horizon, we get the T-dual background (denoted from hereon by a tilde) is,
\begin{eqnarray}
d\tilde{s}^2&=&-\,\frac{dt^2}{1\,-\,\frac{1}{r^{D-3}}}\,+\,\frac{dr^2}{1\,-\,\frac{1}{r^{D-3}}}\,+\,r^2\,d\Omega_{D-2}~,\\
\tilde{\phi}(r)&=&-\,\frac{1}{2}\,\ln(1\,-\,\frac{1}{r^{D-3}})\nonumber~.
\end{eqnarray}
The above background solves (\ref{eq1action}) and (\ref{eq2action}). We refer to this background as the {``\it time-dual of the Schwarzschild background''}. The singularities of the time-dual of the Schwarzschild metric can be seen in the following scalar curvatures, 
\begin{eqnarray}
\tilde{R}&=&-\,\frac{(D\,-\,3)^2}{r^{D-1}\,(r^{D-3}\,-\,1)}~,\\
\tilde{R}_{\mu\nu\lambda\eta}\tilde{R}^{\mu\nu\lambda\eta}&=&\frac{D-3}{r^{2D-2}}\left(\frac{-2(D-2)(3D-7)r^{D-3}+(D-1)\{2D-5+(D-2)^2 r^{2D-6}\}}{(r^{D-3}-1)^2}\right)\nonumber~.
\end{eqnarray} 
Therefore the time-dual of the Schwarzschild metric has two intrinsic singularities at $r=0,1$  which means in particular that the coordinate singularity of the Schwarzschild metric at the horizon has changed to an intrinsic singularity in its time-dual metric. This behaviour is similar to what have been observed in \cite{Tseylinduality} and in \cite{fernando}.

In order to get a better understanding of the time-dual of the Schwarzschild background let us write it in the Einstein frame\footnote{The Einstein frame is obtained by ${g}_{\mu\nu}\,\to\, g_{\mu\nu}\, e^{-\frac{4\phi}{D-2}} $}, 
\begin{eqnarray}\label{efshd}
d\tilde{s}_E^2&=&\left(1\,-\,\frac{1}{r^{D-3}}\right)^{\frac{4-D}{D-2}}\,(-\,dt^2\,+\,dr^2)\,+\,r^2\,\left(1\,-\,\frac{1}{r^{D-3}}\right)^{\frac{2}{D-2}}\,d\Omega_{D-2}~,\\
\tilde{\phi}(r)&=&-\,\frac{1}{2}\,\ln(1\,-\,\frac{1}{r^{D-3}})~.\nonumber
\end{eqnarray}
The singularities of the time-dual of the Schwarzschild metric in the Einstein frame are the same as the ones of the string frame because the Ricci scalar of (\ref{efshd}) reads
\begin{equation}
\tilde{R}_E~=~\frac{(D-3)^2}{D-2}\,\frac{1}{r^{2\,D\,-\,4}}\,\left(\frac{r^{D-3}}{r^{D-3}\,-\,1}\right)^{\frac{D}{D-2}}~.
\end{equation}
The time-dual of the Schwarzschild metric in the Einstein frame for $D>4$  describes a geometry with two  singularities and a Newtonian mass proportional to $(4-D)/(D\,-\,2)$. In $D=4$ the time-dual of the Schwarzschild metric is massless and in the Einstein frame reads
\begin{eqnarray}\label{mlbh}
ds^2&=&-\,dt^2\,+\,dr^2\,+\,r\,|r\,-\,1|\,(d\theta^2\,+\,\sin^2\theta\,d\phi^2)~,\\
\phi(r)&=&-\,\frac{1}{2}\,\ln|1\,-\,\frac{1}{r}|\nonumber~.
\end{eqnarray}
 The above metric (\ref{mlbh}), describes a geometry with  a naked singularity at $r=1$ and a vanishing Newtonian mass. In supergravity similar geometries are studied  and named massless black(white) holes\cite{m1,m2,m3,m4}. Here (\ref{mlbh}) represents the corresponding object in the low energy gravitational theory of the bosonic string theory in $D=4$. Other similar objects in higher dimensions within the low energy theory of the bosonic string theory can be found. For example the background provided below
\begin{eqnarray}\label{mlbhd}
ds^2&=&-\,dt^2\,+\,\frac{dr^2}{1\,+\,\left(\frac{q}{r^{D\,-\,3}}\right)^2} \,+\,r^2\,d\Omega_{D-2}~,\\
\phi&=&\pm~\frac{D\,-\,2}{2\sqrt{D-3}}~{\text ArcSinh}(\frac{q}{r^{D\,-\,3}})~,
\end{eqnarray}
solves the leading order equations of motion in the Einstein frame.
\begin{eqnarray}
R_{\mu\nu}~-~\frac{4}{D-2}\,\nabla_{\mu}\phi\nabla_{\nu}\phi&=&0~,\\ 
\Box \phi&=&0~.
\end{eqnarray}
This background (\ref{mlbhd}), represents a geometry with a vanishing Newtonian mass and an intrinsic singularity at $r=0$ as is clear from the corresponding scalar curvature terms    
\begin{eqnarray}
R&=&(D\,-\,2)\,(D\,-\,3)\,\left(\frac{q}{r^{D-3}}\right)^2~,\\
R_{\mu\nu\eta\xi}\,R^{\mu\nu\eta\xi}&=&2\,(D\,-\,2)\,(D\,-\,3)\,(2\,D\,-\,5)\,\left(\frac{q}{r^{D-2}}\right)^4~.
\end{eqnarray}
To ensure that the dilaton is real we must choose $q$ to also be real. We then 
see that there is a naked singularity at $r=0$. 
 
 The massless black holes are stationary and they should not be thought as massless particles but new vacua of the theory. In superstring it turns out that massless black holes play quite important roles \cite{m1}, a modification of these roles is expected to persist to the bosonic string theory.

\subsection{The linear and the quadratic $\alpha'$ corrections to \\the Schwarzschild background in $D=4$}
The Schwarzschild metric in $D=3+1$ reads
\begin{equation}
ds^2~=~-(1-\frac{1}{r})~dt^2~+~\frac{1}{1-\frac{1}{r}}~dr^2~+~r^2~(d\theta^2~+~\sin^2 \theta~d\phi^2)~,
\end{equation}
where its intrinsic singularity  can be seen in the following scalar curvatures
\begin{eqnarray}
R_{\mu\nu\lambda\eta}\,R^{\mu\nu\lambda\eta}&=&\frac{12}{r^6}~,\\
\nabla_\xi R_{\mu\nu\lambda\eta}\,\nabla^\xi R^{\mu\nu\lambda\eta}&=&\frac{180\,(r\,-\,1)}{r^9}~,
\\
 \nabla_{\xi_1}\nabla_{\xi_2} R_{\mu\nu\lambda\eta}\,\nabla^{\xi_1}\nabla^{\xi_2} R^{\mu\nu\lambda\eta}&=&
 \frac{90}{r^{12}}(56\,r^2\,-\,120\,r\,+\,65)~,
\\
 \nabla_{\xi_1}\nabla_{\xi_2}\nabla_{\xi_3} R_{\mu\nu\lambda\eta}\,\nabla^{\xi_1}\nabla^{\xi_2}\nabla^{\xi_3} R^{\mu\nu\lambda\eta}&=&
 \frac{540\,(r\,-\,1)}{r^{15}}\,(420\,r^2\,-\,1000\,r\,+\,609)~,\label{vanishonhorizon1}\\
 \nabla_{\xi_1}\cdots\nabla_{\xi_4} R_{\mu\nu\lambda\eta}\,\nabla^{\xi_1}\cdots\nabla^{\xi_4} R^{\mu\nu\lambda\eta}&=&
 \frac{270}{r^{18}}(55440\,r^4\,-\,259920\,r^3\,+\,457898\,r^2\\
 &&\qquad-\,358522\,r\,+\,105133)~,\nonumber\\\label{vanishonhorizon2}
 \nabla_{\xi_1}\cdots\nabla_{\xi_5} R_{\mu\nu\lambda\eta}\,\nabla^{\xi_1}\cdots\nabla^{\xi_5} R^{\mu\nu\lambda\eta}&=&
 \frac {540\,(r - 1)}{r^{21}}\,\left(2522520\,r^{4} - 12736080\,r^{3}\right.\\
 && \qquad\quad\left.+ 24176940\,r^{2} - 20406448\,r + 6454623\right)~,\nonumber
\end{eqnarray}
The  Schwarzschild metric can be generalized to a perturbative consistent background of the bosonic string theory in the critical dimension by  assuming that
\begin{enumerate}
\item The critical bosonic string theory admits the following background
\begin{eqnarray}\label{alpha2cs}
ds^2&=&-~g_{tt}(r)\,dt^2\,+\,g_{rr}(r)\,dr^2\,+\,g_\Omega(r)\,(d\theta^2+\sin^2 \theta~d\phi^2)+dx^2_\bot~,\\
\phi&=&\phi(r)~,
\end{eqnarray}
where $dx^2_\bot$ and $\phi$ respectively represent the 22-dimensional flat space and the dilaton.
\item Within the above background the metric and the dilaton admit the following perturbative series in $\alpha'$
\begin{eqnarray}
g_{tt}(r)&=&(1-\frac{1}{r})~(1~+~\alpha'~g_{tt}^{(1)}(r)+~\alpha'^2~g_{tt}^{(2)}(r)~+~\cdots)~,\label{ttsc}\\\label{rrsc}
g_{rr}(r)&=&\frac{1}{1-\frac{1}{r}}~(1~+~\alpha'~g_{rr}^{(1)}(r)~+~\alpha'^2~g_{rr}^{(2)}(r)~+~\cdots)~,\\
g_{_\Omega}(r)&=&r^2~,\\
\phi(r)&=&~0~+~\alpha'~\phi^{(1)}(r)~+~\alpha'^2~\phi^{(2)}(r)~+~\cdots~.\label{phisc}
\end{eqnarray}
\end{enumerate} 
Using the $\beta$-functions (\ref{g2loop}),(\ref{phi2loop}) 
gives the general solution for the linear $\alpha'$ corrections
\begin{eqnarray}
\phi^{(1)}(r)&=&-\,\frac{2\,+\,3\,r\,+\,6\,r^2}{12\,r^3}\,-\,(c_3\,+\,\frac{1}{2})\,\ln(1-\frac{1}{r})\,+\,c_1~,\\\nonumber\\
g_{rr}^{(1)}(r)&=&\frac{10\,-\,3\,r\,-\,6\,r^2\,+\,12\,(c_2+4\,c_3)\,r^3}{12\,r^3\,(r\,-\,1)}\,-\,(c_3+\frac{1}{2})\frac{\ln(1-\frac{1}{r})}{r-1}~,\\\nonumber\\
g_{tt}^{(1)}(r)&=&\frac{6\,+\,5\,r\,+\,12\,r^2\,-\,12\,r^3\,-\,12\,(c_2\,+\,2\,c_3)\,r^3\,+\,c_4\,(r^4\,-\,r^3)}{12\,r^3\,(r\,-\,1)}\,+\\
&&+\,(3\,-\,2\,r)\,(c_3\,+\,\frac{1}{2})\,\frac{\ln(1\,-\,\frac{1}{r})}{r\,-\,1}~.\nonumber
\end{eqnarray}
In \cite{callan} and many following works a particular boundary conditions were chosen for the metric and dilaton. In these works it was assumed that after choosing (\ref{ttsc}) and (\ref{rrsc}) as the perturbative series for the metric then the $\alpha'$ corrections to the metric are finite at the horizon of the black hole and the dilaton vanishes at infinity,
\footnote{One star is used upon the metric and dilaton with these boundary conditions. We only review  these metrics and dilaton and we are not going to use them.}
\begin{eqnarray}\label{callanboundary}
\phi^{\star(1)}(r)|_{r=\infty}&=&0~,\\
g_{tt}^{\star(1)}(r)|_{r=1}&<&\infty~,\nonumber\\
g_{rr}^{\star(1)}(r)|_{r=1}&<&\infty~.\nonumber
\end{eqnarray}
The above boundary conditions fix the constants of the integration to values of $c_1=0$, $c_2=\frac{23}{12}$, $c_3=-\frac{1}{2}$, $c_4=0$ giving
\begin{eqnarray}
 g_{tt}^{\star(1)}(r)&=&-\,\frac{23\,r^2\,+\,11\,r\,+\,6}{12\,r^3}~,\\
g_{rr}^{\star(1)}(r)&=&-\,\frac{r^2\,+\,7\,r\,+\,10}{12\,r^3}~,\\
\phi^{\star(1)}(r)&=&-\,\frac{2\,+\,3\,r\,+\,6\,r^2}{12\,r^3}~.
\end{eqnarray}
The boundary conditions imposed in (\ref{callanboundary}) produce finite corrections to the Hawking temperature and the entropy of the black-hole. In addition it produces  corrections to the Newtonian mass which is provided by the asymptotic behaviour of $g_{tt}(r)$ at large $r$.

In the next section we are going first to calculate the $\alpha'$ corrections to the time-dual of the Schwarzschild metric and then we will use the $\alpha'$ modified T-duality rules.  We obtained these rules  by studying the $\alpha'$ corrections to the Kasner background. Within the $\alpha'$-calculations we fixed the asymptotic behaviours at infinity both for the Kasner background and its dual. To be consistent with those calculations and due to the intrinsic singularity of the time-dual of the Schwarzschild metric we use the following boundary conditions on the field contents of the Schwarzschild background (as opposed to the
boundary condition in (\ref{callanboundary})),
\begin{eqnarray}
\phi(r)|_{r=\infty}&=&0~,\\\nonumber
g_{tt}(r)|_{r\sim\infty}&=&1\,-\,\frac{1}{r}\,+\,O(\frac{1}{r^2})~,\\\nonumber
g_{rr}(r)|_{r\sim\infty}&=&1\,+\,\frac{1}{r}\,+\,O(\frac{1}{r^2})~.
\end{eqnarray}
Choosing the above boundary conditions and using the $\beta$ functions (\ref{g2loop}),(\ref{phi2loop}) identifies the linear $\alpha'$ corrections 
\begin{eqnarray}\label{asmbaound}
g_{tt}^{(1)}(r)&=&\frac{6\,+\,5\,r\,+\,12\,r^2\,-\,12\,r^3\,+(18\,r^3\,-12\,r^4)\,\ln(1-\frac{1}{r})}{12~r^3~(r\,-\,1)}~,\\
g_{rr}^{(1)}(r)&=&\frac{10\,-\,3\,r\,-\,6\,r^2\,-6\,r^3\,\ln(1-\frac{1}{r})}{12~r^3~(r\,-\,1)}~,\nonumber\\
\phi^{(1)}(r)&=&-\,\frac{1\,+\frac{3}{2}\,r\,+\,3\,r^2\,+\,3\,r^3\,\ln(1-\frac{1}{r})}{6~r^3}~.\nonumber
\end{eqnarray}
as well as the quadratic $\alpha'$ corrections
\begin{eqnarray}\label{alphacorrectiondiverge1}
g_{tt}^{(2)}(r)&=&\frac{7050-5758 r + 8125 r^2 - 2757 r^3 + 1940 r^4 + 8140 r^5 - 19680 r^6 + 6240 r^7}{7200\, r^6\,(r\,-\,1)^2}\nonumber\\
&&~+~\ln(1-\frac{1}{r})~\frac{120~-~45~r~+~45~r^2~+~126~r^3~-~320~r^4~+~104~r^5}{120~r^3~ (r~-~1)^2}\nonumber\\
&&~+~\{\ln(1-\frac{1}{r})\}^2~\frac{4~r~- ~ 9}{8~ (r~-~1)^2}~,\\
\nonumber\\\nonumber\\\label{alphacorrectiondiverge2}
g_{rr}^{(2)}(r)&=&\frac{6250-\,10154\,r\,+\,5049\,r^2\,-\,5135\,r^3\,+1880\,r^4-2460\,r^5\,-480\,r^6}{7200~r^6~(r-1)^2}\nonumber\\
&&-\,\ln(1-\frac{1}{r})\,\frac{85\,-\, 75\, r\, + 22\, r^2\, + 8\, r^3}{120~r^2~(r-1)^2}~+~\{\ln(1-\frac{1}{r})\}^2~\frac{1~+~r}{8~(r\,-\,1)^2}~,
\\\nonumber\\\nonumber\\\label{alphacorrectiondiverge3}
\phi^{(2)}(r)&=&\frac{-225\,-\,327\,r\,-513\,r^2-\,205\,r^3-710\,r^4\,-3930\,r^5\,+4260\,r^6\,}{3600~r^6~(r-1)}\nonumber\\
&&+~\ln(1\,-\,\frac{1}{r})~\frac{-\,5\,+\,5\,r\,+\,15\,r^2\,-\,101\,r^3\,+\,71\,r^4}{60~r^3~(r\,-\,1)}~.
\end{eqnarray}
Inserting the above linear and quadratic $\alpha'$ corrections in (\ref{ttsc},\ref{rrsc},\ref{phisc}) identifies the quadratic $\alpha'$ corrected Schwarzschild background.  The asymptotic behaviours of the fields of the quadratic $\alpha'$ corrected Schwarzschild background at large r follow
\begin{eqnarray}
g_{rr}(r)\,(1\,-\,\frac{1}{r})&=&1\,+\,\frac{40\,r^2\,+\,45\,r\,+\,49}{40\,r^6}\,\alpha'\,+\,\frac{9}{8\,r^6}\alpha'^2\,+\,O(\frac{1}{r^7})~,\\
g_{tt}(r)~\frac{1}{1\,-\,\frac{1}{r}}&=&1\,+\,\frac{30\,r^2\,+\,9\,r\,-\,7}{120\,r^6}\,\alpha'\,+\,\frac{3}{8\,r^6}\alpha'^2\,+\,O(\frac{1}{r^7})~,\\
\phi(r)&=&\frac{105\,r^3\,+\,84\,r^2\,+\,70\,r\,+\,60}{840\,r^7}\,\alpha'\,+\,\frac{1}{168\,r^7}\,\alpha'^2\,+\,O(\frac{1}{r^8}). 
\end{eqnarray}

Looking carefully at these expressions one notices that at an $\alpha'$ 
dependent location outside what was the horizon at $r=1$ the component 
$g_{tt}$ of the metric passes through zero. At this point both $g_{rr}$ and
$\phi(r)$ remain finite. If the zero in $g_{tt}$ happens at $r=r_0$ then 
defining a new coordinate $\rho = r - r_0$ the metric near this zero 
takes the form, 
\begin{equation}
ds^2 = -\rho dt^2 + d\rho^2 + c^2 d\omega^2,
\end{equation}
where $c$ is a real constant. This metric has a singular Ricci scalar 
at $\rho=0$ and thus the $\alpha'$ corrections to the Schwarzschild metric
appear to give rise to a naked singularity. The fact that the generic 
$\alpha'$ corrections to the Schwarzschild metric have singularities 
outside what was the horizon at $r=1$ was already noted in \cite{callan}.

\subsection{The linear and the quadratic $\alpha'$ corrections to the time-dual of the Schwarzschild background in $D=4$}
The time dual of the Schwarzschild metric in $D=4$ is
\begin{eqnarray}
ds^2&=&-~\frac{1}{1-\frac{1}{r}}~dt^2~+~\frac{1}{1-\frac{1}{r}}~dr^2~+~r^2~(d\theta^2~+~\sin^2 \theta~d\phi^2)~,\\
\phi&=&-\frac{1}{2}\ln(1-\frac{1}{r})~.
\end{eqnarray}
 The various scalar curvatures show intrinsic singularities at $r=0$ and at $r=1$, 
\begin{eqnarray}
R&=&\frac{1}{r^2~(-r~+~1)}~,\\
R_{\mu\nu}R^{\mu\nu}&=&\frac{9~-20~r~+~12~r^2}{2~r^6~(-~r~+~1)^2}~,\\
R_{\mu\nu\eta\xi}R^{\mu\nu\eta\xi}&=&\frac{9~-20~r~+~12~r^2}{r^6~(-~r~+~1)^2}~,\\\label{csdDRDR} 
\bigtriangledown_\gamma R_{\mu\nu\eta\xi}\bigtriangledown^\gamma R^{\mu\nu\eta\xi}&=&\frac{180~r^4~-~648~r^3~+~900~r^2~-~568~r~+~137\underline{}}{(r~-~1)^3~r^9}~.
\end{eqnarray}
The local string coupling constant in this background reads
\begin{equation}
g_s~=~e^{+\,\phi}~=~\frac{g_0}{\sqrt{1\,-\,\frac{1}{r}}}~.
\end{equation}
Therefore far away from $r=1$ both the string theory is perturbative and the space-time is asymptotically flat. Thus within this regime the time-dual of the Schwarzschild metric  can be extended to the following perturbative background in the critical bosonic string theory
\begin{eqnarray}\label{dalpha2cs}
ds^2&=&-~ G_{tt}(r)~dt^2~+~G_{rr}(r)~dr^2~+~G_{\Omega}(r)~(d\theta^2~+~\sin^2 \theta~d\phi^2)~+~dx^2_\bot~,\nonumber\\
\phi&=& \Phi(r)~,\nonumber
\end{eqnarray}
where the metric and the dilaton admit the following perturbative series in $\alpha'$
\begin{eqnarray}
G_{tt}(r)&=&\frac{1}{1-\frac{1}{r}}~(1~+~\alpha'~G_{tt}^{(1)}(r)+~\alpha'^2~G_{tt}^{(2)}(r)~+~\cdots)~,\\
G_{rr}(r)&=&\frac{1}{1-\frac{1}{r}}~(1~+~\alpha'~G_{rr}^{(1)}(1)~+~\alpha'^2~G_{rr}^{(2)}(r)~+~\cdots)~,\\
G_{\Omega}(r)&=&r^2~,\\
\Phi(r)&=&~-\frac{1}{2}\ln(1-\frac{1}{r})~+~\alpha'~\Phi^{(1)}(r)~+~\alpha'^2~\Phi^{(2)}(r)~+~\cdots~.
\end{eqnarray}
We fix the asymptotic behaviours of the metric and the dilaton at large $r$ by
\begin{eqnarray}
\Phi(r)|_{r\sim\infty}&=&\frac{1}{2}\,\frac{1}{r}\,+\,O(\frac{1}{r^2})~,\\
G_{tt}(r)|_{r\sim\infty}&=&1\,+\,\frac{1}{r}\,+\,O(\frac{1}{r^2})~,\\
G_{rr}(r)|_{r\sim\infty}&=&1\,+\,\frac{1}{r}\,+\,O(\frac{1}{r^2})~.
\end{eqnarray} 
{In \cite{confusing} the linear $\alpha'$ corrected Schwarzschild metric computed in \cite{callan} with the boundary condition provided in (\ref{callanboundary}) and the linear $\alpha'$ modified T-duality is used to obtain the linear $\alpha'$ corrections to the time-dual of the Schwarzschild metric in $D=4$. This procedure means choosing a specific boundary condition for the time-dual of the Schwarzschild metric at $r=1$. However we think that due to the intrinsic singularity at $r=1$ it is not reasonable to set any boundary condition at $r=1$ in the time-dual of the Schwarzschild background.}

 Using the $\beta$-functions  (\ref{g2loop}), (\ref{phi2loop}) and  the above asymptotic behaviours  identifies the linear $\alpha'$ corrections
\begin{eqnarray}  
G_{tt}^{(1)}(r)&=&\frac{\frac{5}{4}~+~3~r~-~3~r^2~-~(3~r^3~-~\frac{9}{2}~r^2)~\ln(1~-~\frac{1}{r})}{3~r^2~(-~r~+~1)}~,\\
G_{rr}^{(1)}(r)&=&
\frac{10\,-\,3\,r\,-\,6\,r^2\,-6\,r^3\,\ln(1-\frac{1}{r})}{12~r^3~(r\,-\,1)}~,\\
\Phi^{(1)}(r)&=&\frac{1~-~3~r~-~6~r^2~-~6~r^3~\ln(1-\frac{1}{r})}{24~r^3~(r~-~1)}~,
\end{eqnarray} 
as well as the quadratic $\alpha'$ corrections
\begin{eqnarray}\nonumber
G_{tt}^{(2)}(r)&=&\frac{300 - 2042 \,r + 2125 \,r^2 + 1557\, r^3- 740\, r^4- 22540\,r^5+ 26880\,r^6  - 6240\,r^7 }{7200~r^6~(r~-~1)^2}\,+\nonumber\\
&&+~\ln(1-\frac{1}{r})~\frac{75+215 r-726 r^2+560 r^3-104 r^4}{120~r^2~(r~-~1)^2}\\
&&+~\{\ln(1-\frac{1}{r})\}^2~\frac{9-11r+4r^2}{8~(r-1)^2}~,\nonumber\\\nonumber
\\\nonumber\\\nonumber
G_{rr}^{(2)}(r)&=&\frac{2200-4754\,r+\,5049\,r^2-5135\,r^3+1880\,r^4-2460\,r^5-480\,r^6}{7200~r^6~(r~-~1)^2}\\
&&+~\ln(1-\frac{1}{r})~\frac{-85+75 r -22 r^2- 8r^3}{120~r^2~(r-1)^2}~+~\{\ln(1-\frac{1}{r})\}^2~\frac{r+1}{8~(r-1)^2}~,\\
\nonumber\\\nonumber\\\nonumber
\Phi^{(2)}(r)&=&\frac{-1350+2116\,r-456\,r^2-875\,r^3+680\,r^4-2460\,r^5-9480\,r^6+10800\,r^7}{14400\,r^{6}\,(r - 1)^{2}}\\
&&+\ln(1-\frac{1}{r})\,\frac {5  - 10\,r + 45\,r^{2} + 38\,r^{3} - 248\,r^{4} + 180\,r^{5} }{240\,r^{3}\,(r - 1)^{2}}\\
&&+\,\{\ln(1-\frac{1}{r})\}^2\frac{r}{16(r\,-\,1)^2}~.\nonumber
\end{eqnarray}

\subsection{Applying the quadratic $\alpha'$ modified T-duality on the $\alpha'^2$ corrected Schwarzschild background and its dual in $D=4$}

Earlier we obtained the $\alpha'$ modified T-duality rules for time-dependent geometries. These same rules should also describe T-duality in the Schwarzschild
metric, but now the T-duality acts in the direction of the time-like 
Killing vector outside the black hole horizon. These  include the Euclidean geometry of the Schwarzschild metric. Consequently  the $\alpha'$ modified T-duality rules can be legitimately  applied to the Schwarzschild  background. In order to do this we  first introduce the analog of the co-moving and the almost co-moving frame for the metrics of the Schwarzschild background and its time-dual. 

The quadratic $\alpha'$ corrected Schwarzschild metric and its time-dual are spherically symmetric. A spherically symmetric metric can be written in a coordinate where the radial component of the metric is the identity. This coordinate is the analog of the co-moving frame which we are going to refer to  as the {``\it radial co-moving frame''}. On the other hand  the {``\it almost radial co-moving frame''} is defined as a coordinate which can be transformed to the radial co-moving frame by a {\it single} re-parametrization of the radial direction . In the almost radial co-moving frame the metric reads 
\begin{equation}
ds^2~=~f(r)~dr^2~+~ds_\bot^2~,
\end{equation}
where $f(r)$ is called the {``\it characteristic function''} of the almost radial co-moving frame. The quadratic $\alpha'$ Schwarzschild metric in (\ref{alpha2cs}) and its time-dual in (\ref{dalpha2cs}) are already  in the almost radial co-moving frame respectively with $g_{rr}(r)$ and $G_{rr}(r)$ as their characteristic functions. These two characteristic functions are not equal: 
\begin{equation}
G_{rr}(r)~-~g_{rr}(r)~=~\frac{3\,(4\,r\,-\,3)~\alpha'^2}{16~(r\,-\,1)^3\,r^5}~,
\end{equation}
In order to apply the $\alpha'$ modified T-duality rules to the $\alpha'$ corrected Schwarzschild metric and its $\alpha'$ corrected time-dual we  should first write them in the almost radial co-moving frame with the same characteristic functions. In the following, we  write 
 the quadratic $\alpha'$ corrected time-dual of the Schwarzschild metric in the almost radial co-moving frame with  the characteristic function of the  quadratic $\alpha'$ corrected Schwarzschild metric.

The $\alpha'$ corrected time-dual of the Schwarzschild metric in the new coordinate provided by  
\begin{eqnarray}
r&\to&r~-~\frac{\alpha'^2}{16~(r\,-\,1)~r^5}~,
\end{eqnarray}
reads
\begin{eqnarray}\label{alpha2dualcs}
ds^2&=&-~\tilde{g}_{tt}(r)~dt^2~+~g_{rr}(r)~dr^2~+~\tilde{g}_{_\Omega}(r)~(d\theta^2\,+\,\sin^2\theta\,d\phi^2)~,\\
\phi&=&\tilde{\phi}(r)~,\nonumber
\end{eqnarray}
where $g_{rr}(r)$ is given by (\ref{rrsc}). Other components of the metric and the dilaton read
\begin{eqnarray}
\tilde{g}_{tt}(r)&=&\frac{1}{1-\frac{1}{r}}~(1~+~\alpha'~\tilde{g}_{tt}^{(1)}(r)+~\alpha'^2~\tilde{g}_{tt}^{(2)}(r))~,\\
\tilde{g}_{_\Omega}(r)&=&r^2~+~\frac{\alpha'^2}{8~r^4~(r-1)}~,\\
\tilde \phi(r)&=&-\,\frac{1}{2}\ln(1-\frac{1}{r})~+~\alpha'~\tilde{\phi}^{(1)}(r)~+~\alpha'^2~\tilde{\phi}^{(2)}(r)~.
\end{eqnarray}
The $\alpha'$ coefficients to the dilaton $\phi(r)$ are
\begin{eqnarray}
\tilde{\phi}^{(1)}(r)&=&\frac{1\,-\,3\,r\,-\,6\,r^2\,-\,6\,r^3\,\ln(1-\frac{1}{r})}{24\,r^3\,(r\,-\,1)}~,\\
\nonumber\\\nonumber
\tilde{\phi}^{(2)}(r)&=& \frac {- 1800 + 2116 r - 456 r^{2} + 
10800 r^{7} - 875 r^{3} + 680 r^{4} - 2460 r^{5} - 9480 r^{6}}{14400\,r^{6}\,(r - 1)^{2}}\\
&&+ \ln(1-\frac{1}{r}) \,{ \frac { - 248\,r^{4} + 180\,r^{5} + 38\,r^{3} - 10\,r + 45\,r^{2} + 5}{240~r^{3}\,(r - 1)^{2}}}\\
&&+\{\ln(1-\frac{1}{r})\}^2\frac{r}{16\,(r-1)^2}~.\nonumber
\end{eqnarray}
The $\alpha'$ coefficients to the $\tilde{g}_{tt}(r)$ read
\begin{eqnarray}
\tilde{g}_{tt}^{(1)}(r)&=&\frac{-\,5\,-\,12\,r\,+\,12\,r^2\,+\,(12\,r^3\,-\,18\,r^2)\,\ln(1\,-\,\frac{1}{r})}{12\,r^2\,(r\,-\,1)}~,\\
\tilde{g}_{tt}^{(2)}(r)&=&-\frac {150  + 2042\,r - 2125\,r^{2} - 1557\,r^{3} + 740\,r^{4} + 22540\,r^{5} - 26880\,r^{6} + 6240\,r^{7}}
{7200\,r^{5}\,(r - 1)^{3}}\nonumber\\
&&+\ln(1-\frac{1}{r})~\frac {215\,r + 75 - 104\,r^{4} - 726\,r^{2}+ 560\,r^{3}}{120\,r\,(r - 1)^{3}}\\
&&+\{\ln(1-\frac{1}{r})\}^2\frac {r\,( - 11\,r + 9 + 4\,r^{2})}{8\,(r - 1)^{3}}~.\nonumber 
\end{eqnarray}
This new coordinate is the almost radial co-moving frame with the characteristic function $g_{rr}(r)$. Now the $\alpha'$ modified T-duality rules provided by (\ref{tda1},\ref{tda2},\ref{tda3}) can be applied to the quadratic $\alpha'$ corrected Schwarzschild background (\ref{alpha2cs}) and its time-dual (\ref{alpha2dualcs}).

The $tt$, $\phi\phi$ and $\theta\theta$ components of the quadratic $\alpha'$ corrected Schwarzschild background (\ref{alpha2cs}) and its time dual (\ref{alpha2dualcs}) satisfy  the T-duality rules given by (\ref{tda1}) and (\ref{tda2}) 
\begin{eqnarray}
\label{gtt_check_sc}
\ln \tilde g_{tt}&-&\frac{\alpha'}{4}~\tilde{\nabla}_\mu\ln\tilde{g}_{tt}\,.\,\tilde{\nabla}^\mu\ln\tilde{g}_{tt}\\
 &=&-\,\{\ln g_{tt}~-~\frac{\alpha'}{4}~{\nabla}_\mu\ln{g}_{tt}\,.\,{\nabla}^\mu\ln{g}_{tt}\}~+~O(\alpha'^2\nabla^4)~,\nonumber\\
\label{grr_check_sc}
\ln \tilde{g}_{_{ii}}&+&\frac{\alpha'^2}{32}\tilde{\nabla}_\mu\ln \tilde{g}_{tt}.\tilde{\nabla}^\mu\ln \tilde{g}_{_{ii}}~\tilde{\nabla}_\nu\ln \tilde{g}_{tt}.\tilde{\nabla}^\nu\ln \tilde{g}_{tt}\\
&=&\ln {g}_{_{ii}}+\frac{\alpha'^2}{32}{\nabla}_\mu\ln {g}_{_{ii}}.{\nabla}^\mu\ln {g}_{tt}~{\nabla}_\nu\ln
{g}_{tt}.{\nabla}^\nu\ln {g}_{tt}+O(\alpha'^3\nabla^6)~.\nonumber
\end{eqnarray}
Note that in (\ref{tda3}) we should substitute $\det g$ and $\det \tilde g$ respectively with ${\det}^*g$ and ${\det}^*\tilde g$ given by 
\begin{eqnarray}
{det}^* g&=&\frac{\det g}{g_{_{rr}}(r)}~,\\
{det}^* \tilde g&=&\frac{\det \tilde g}{g_{_{rr}}(r)}~.
\end{eqnarray}
In order to check (\ref{tda3}) we first calculate 
\begin{equation}
\tilde\phi(r)\,-\,\frac{1}{4}\,\ln\det\tilde g\,-\phi(r)\,+\,\frac{1}{4}\,\ln\det g~=~\frac{3\,-\,4\,r}{32~r^6~(r\,-\,1)^2}\,\alpha'^2~.
\end{equation}
Also the $\alpha'^2$ terms in the l.h.s of (\ref{tda3})
\begin{eqnarray}
\tilde{\nabla}_\mu\ln \tilde{g}_{_{tt}}~\tilde{\nabla}_\nu\ln \tilde{g}_{_{t\,t}}~\tilde{\nabla}^\mu\tilde{\nabla}^\nu\ln \tilde{g}_{_{t\,t}}&=&\frac{4\,r\,-\,3}{2~r^6~(r\,-\,1)^2}\,+\,O(\alpha')~,\\
\label{futureformula1}
\tilde{\nabla}_\mu\ln \tilde{g}_{_{t\,t}}.\tilde{\nabla}^\mu\ln \tilde{g}_{_{t\,t}}~\tilde{\nabla}_\nu\ln \tilde{g}_{_{t\,t}}.\tilde{\nabla}^\nu({\tilde{\phi}}-\frac{1}{4}\ln {\det}^*\tilde g)&=&\frac{4\,r\,-\,5}{4\,r^6~(r\,-\,1)^2}\,+\,O(\alpha')~,\\
\tilde{\nabla}_\mu \ln \tilde{g}_{_{t\,t}}.\tilde{\nabla}^\mu\ln \tilde{g}_{_{t\,t}}~\tilde{\Box}\ln \tilde{g}_{_{t\,t}}&=&\frac{1}{r^6~(r\,-\,1)^2}\,+\,O(\alpha')~,
\end{eqnarray}
and the $\alpha'^2$ terms in the r.h.s of (\ref{tda3})
\begin{eqnarray}
\nabla_\mu\ln g_{_{t\,t}}~\nabla_\nu\ln g_{_{t\,t}}~\nabla^\mu\nabla^\nu\ln g_{_{t\,t}}&=&-\,\frac{4\,r\,-\,3}{2~r^6~(r\,-\,1)^2}\,+\,O(\alpha')~,\\
\label{futureformula2}
\nabla_\mu\ln g_{_{t\,t}}.\nabla^\mu\ln g_{_{t\,t}}~\nabla_\nu\ln g_{_{t\,t}}.\nabla^\nu(\phi-\frac{1}{4}\ln {\det}^* g)&=&-\,\frac{4\,r\,-\,3}{4\,r^6~(r\,-\,1)^2}\,+\,O(\alpha')~,\\
\nabla_\mu \ln g_{_{t\,t}}.\nabla^\mu\ln g_{_{t\,t}}~\small{\Box}\ln g_{_{t\,t}}&=&0~+~O(\alpha')~.
\end{eqnarray}
Inserting the above  expressions in (\ref{tda3}) results  the following equations for $A$, $B$ and $C$
\begin{eqnarray}
\left.
\begin{array}{r c r}
A\,+\,B\,+\,C&=&1\\
A\,+\,B&=&1\\
3\,A\,+\,4\,B\,-\,C&=&3
\end{array}
\right\}
\Longrightarrow
\left\{
\begin{array}{r c r}
A&=&1\\
B&=&0\\
C&=&0
\end{array}
\right.
\end{eqnarray}
which identifies $A=1$ and $B=C=0$. Substituting these values in (\ref{tda3}) results 
\begin{eqnarray}\label{tda3fixed}
\tilde{\phi}-\frac{1}{4}\ln \det\tilde g&+&\frac{\alpha'^2}{32}~\tilde{\nabla}_\mu\ln \tilde{g}_{_{25\,25}}~\tilde{\nabla}_\nu\ln \tilde{g}_{_{25\,25}}~\tilde{\nabla}^\mu\tilde{\nabla}^\nu\ln \tilde{g}_{_{25\,25}}\\
&=&\phi-\frac{1}{4}\ln \det g\,+\,\frac{\alpha'^2}{32}~\nabla_\mu\ln g_{_{25\,25}}~\nabla_\nu\ln g_{_{25\,25}}~\nabla^\mu\nabla^\nu\ln g_{_{25\,25}}~+~O(\alpha'^3\nabla^6)\nonumber
\end{eqnarray}
The fact that we can find a consistent assignment of the constants $A$, $B$, $C$ and satisfy (\ref{gtt_check_sc}) and (\ref{grr_check_sc})  is a nontrivial check on the consistency of our procedure.

As mentioned earlier a general proof of the validity of T-duality at up to 
third order in $\alpha'$ is presented in \cite{odd}. This implies that 
once we have fully fixed the constants $A, B$ and $C$, there is no need 
to check further that this form is consistent with the T-duality for 
other metrics. One may of course argue that applying T-duality to the 
Schwarzschild metric is not completely conventional. To resolve this 
possible ambiguity one could follow the above algorithm applied 
to another homogeneous cosmology for example of the type studied in
\cite{kkunz} where T-duality and cosmology is studied in some detail.

\section{Conclusion}

In this work the linear and the quadratic $\alpha'$-corrections to the general diagonal Kasner background in the critical bosonic string theory at the tree level of string interactions are computed. 

The asymptotic behaviours of the dilaton and the metric are fixed at {\it infinity} and  the linear and the quadratic $\alpha'$ corrections to the Schwarzschild metric in $D=4$ are computed. Regardless of the mass of the black hole these corrections diverge at the horizon.

By applying T-duality in the time direction of the  
 Schwarzschild metric the {``\it time-dual of the Schwarzschild metric''} is introduced. 
It is observed that the coordinate singularity of the Schwarzschild metric at the horizon changes into an intrinsic singularity in its T-dual metric.

We find it quite curious that the consistency of $\alpha'$ corrections 
and T-duality leads us to the conclusion that the 
$\alpha'$ corrected Schwarzschild metric with singular horizon plays
a more important role than that fine tuned to have a smooth horizon. 
To understand the possible relevance of this result one would need
to study string theory on such backgrounds to verify the actual presence
of T-duality - equivalence of physics between T-dual geometries under
the interchange of momentum and winding modes.

The asymptotic behaviours of the fields of the time-dual of the Schwarzschild metric are fixed at infinity and the linear and the quadratic $\alpha'$ corrections to the time-dual of the Schwarzschild metric in $D=4$ are computed.   

The time-dual of the Schwarzschild metric is massless in $D=4$. As a generalization of this geometry we introduced the following background
\begin{eqnarray}\label{masslessblackhole}
ds^2&=&-\,dt^2\,+\,\frac{dr^2}{1\,+\,\left(\frac{q}{r^{D\,-\,3}}\right)^2} \,+\,r^2\,d\Omega_{D-2}~,\\
\phi&=&\pm~\frac{D\,-\,2}{2\sqrt{D-3}}~{\text ArcSinh}(\frac{q}{r^{D\,-\,3}})~.\nonumber
\end{eqnarray}
The above background represent a non-trivial (and singular) massless geometry in an arbitrary dimension for the low energy gravitational theory of the bosonic string.

The quadratic $\alpha'$ modification to the rules of T-duality for time dependent backgrounds composed of dilaton and diagonal metric are obtained. If we choose the co-moving frame to represent the dilaton and the metric 
\begin{eqnarray}\label{timedependentbk}
ds^2&=&-\,dt^2\,+\,\sum_{i=1}^{25}g_{ii}(t)\,dx^2_i~,\\
\phi&=&\phi(t)\nonumber~,
\end{eqnarray}
and  choose the co-moving frame in the T-dual space to represent the T-dual dilaton and metric 
\begin{eqnarray}\label{dtimedependentbk}
d\tilde{s}^2&=&-\,dt^2\,+\,\sum_{i=1}^{25}\tilde{g}_{ii}(t)\,dx^2_i~,\\
\tilde\phi&=&\tilde\phi(t)\nonumber~,
\end{eqnarray}
then in  the string frame the leading  $\alpha'$ modification to rules  describing T-duality in $x_{_{25}}$ direction and relating (\ref{timedependentbk}) to (\ref{dtimedependentbk}) when the $\beta$ functions ,(\ref{g2loop}) and (\ref{phi2loop}), are calculated by the dimensional regularization method in the minimal substraction scheme  read
\begin{eqnarray}\label{ftda1}
\ln \tilde{g}_{_{25\,25}}&-&\frac{\alpha'}{4}\tilde{\nabla}_\mu\ln \tilde{g}_{_{25\,25}}.\tilde{\nabla}^\mu\ln \tilde{g}_{_{25\,25}}\\\nonumber\label{ftda2}
&=&-\left\{\ln {g}_{_{25\,25}}-\frac{\alpha'}{4}{\nabla}_\mu\ln {g}_{_{25\,25}}.{\nabla}^\mu\ln {g}_{_{25\,25}}\right\}+O(\alpha'^2\nabla^4)~,\\\nonumber\\
\ln \tilde{g}_{_{ii}}&+&\frac{\alpha'^2}{32}\tilde{\nabla}_\mu\ln \tilde{g}_{_{25\,25}}.\tilde{\nabla}^\mu\ln \tilde{g}_{_{ii}}~\tilde{\nabla}_\nu\ln \tilde{g}_{_{25\,25}}.\tilde{\nabla}^\nu\ln \tilde{g}_{_{25\,25}}\\
&=&\ln \tilde{g}_{_{ii}}+\frac{\alpha'^2}{32}{\nabla}_\mu\ln {g}_{_{ii}}.{\nabla}^\mu\ln {g}_{_{25\,25}}~{\nabla}_\nu\ln {g}_{_{25\,25}}.{\nabla}^\nu\ln {g}_{_{25\,25}}+O(\alpha'^3\nabla^6)\nonumber~,
\\\nonumber\\
\tilde{\phi}&-&\frac{1}{4}\ln \det\tilde g\,+\,\frac{\alpha'^2}{32}~\tilde{\nabla}_\mu\ln \tilde{g}_{_{25\,25}}~\tilde{\nabla}_\nu\ln \tilde{g}_{_{25\,25}}~\tilde{\nabla}^\mu\tilde{\nabla}^\nu\ln \tilde{g}_{_{25\,25}}\\
&=&\phi\,-\,\frac{1}{4}\ln \det g\,+\,\frac{\alpha'^2}{32}~\nabla_\mu\ln g_{_{25\,25}}~\nabla_\nu\ln g_{_{25\,25}}~\nabla^\mu\nabla^\nu\ln g_{_{25\,25}}~+~O(\alpha'^3\nabla^6)\nonumber~.
\end{eqnarray}
As mentioned before within the convention provided by the $\beta$ functions, (\ref{ftda2}) means that applying T-duality in a general time-dependent background alters the metric in all directions.
These rules are consistent with the linear $\alpha'$ corrections
to T-duality studied in \cite{Tseylinduality,olsen}, and have also been 
checked to agree with T-duality and quadratic $\alpha'$ corrections for the 
Schwarzschild
metric in $D=5$ space-time dimensions.

In case of the Kasner background, keeping the quadratic $\alpha'$ terms in (\ref{ftda1}) results in 
\begin{eqnarray}\label{tda12}
\ln \tilde{g}_{_{25\,25}}&-&\frac{\alpha'}{4}\,\tilde{\nabla}_\mu \ln \tilde{g}_{_{25\,25}}\tilde{\nabla}^\mu \ln\tilde{g}_{_{25\,25}}\,+\, \frac{\alpha'^2}{32}(\tilde{\nabla}_\mu \ln \tilde{g}_{_{25\,25}}\tilde{\nabla}^\mu \ln\tilde{g}_{_{25\,25}})^2\,-\,\frac{\alpha'^2\,\tilde{p}_{_{25}}^2}{2\,t^4}\\
&=&-\left\{\ln g_{_{25\,25}}-\frac{\alpha'}{4}\nabla_\mu \ln g_{_{25\,25}}\nabla^\mu\ln g_{_{25\,25}}
+ \frac{\alpha'^2}{32}({\nabla}_\mu \ln {g}_{_{25\,25}}{\nabla}^\mu \ln{g}_{_{25\,25}})^2-\frac{\alpha'^2{p}_{_{25}}^2}{2t^4}\right\}\nonumber
\end{eqnarray}
There exist nine possibilities to write $\frac{p_{_{25}}^2}{2\,t^4}$ as derivatives of the dilaton and logarithm of the Kasner metric.
Using the quadratic $\alpha'$ corrected Schwarzschild metric  and its T-dual in $D=4$ and $D=5$ fails to single out a unique possibility. 
In order to write (\ref{tda12}) unambiguously for a general time-dependent background one should study the quadratic $\alpha'$ corrections to other backgrounds such as the Schwarzschild metric in higher dimensions, the massless black holes provided in (\ref{masslessblackhole}) or the homogeneous cosmologies 
studied in \cite{kkunz} and their corresponding T-duals.

As a generalization of this work one may consider the four-loop $\alpha'$ corrections in the critical bosonic string theory where the corresponding  $\beta$-functions are computed in \cite{Jack2} or the four-loop $\alpha'$  corrections in superstring theory \cite{s1,s2,s3,s4,s5,s6}.    

\section{Acknowledgments}
The authors would like to thank M. Blau for discussions and for a critical reading of the almost final version of this paper. 
One of the authors G. Exirifard wishes to thank L. Bonora for comments and discussions throughout this work. He also wishes to thank J. David, M.H. Sarmadi and M. Serone for useful discussions. In addition he appreciates his office-mates F. Alday, M. Cirafici and C. Maccaferri for the set of long consecutive discussions. Finally he would like to thank  H.S. Allaei and C. Attaccalite for their help in installing linux and needed softwares on his laptop and E. Arabzadeh for his suggestions on the English structure of the  early draft. This work is supported in part by the European Community's Human Potential
Programme under contract HPRN-CT-2000-00131 Quantum Spacetime.

\end{document}